\documentclass[twocolumn, aps,pra,superscriptaddress,showpacs,floatfix]{revtex4-1}

\UseRawInputEncoding
\usepackage{float}
\usepackage{dcolumn}
\usepackage{bm}
\usepackage[colorlinks = true, linkcolor = blue, urlcolor  = blue, citecolor = blue, anchorcolor = blue]{hyperref}
\usepackage{longtable}
\usepackage{booktabs}
\usepackage{mathrsfs}
\usepackage{graphicx,epsfig,latexsym,amssymb}
\usepackage{multirow,amsmath,array,booktabs,color}
\usepackage[section]{placeins}
\usepackage{color}
\usepackage{soul}

\begin{document}

\title{The process of superradiant phase transition for quantum Rabi model in view of  nonclassical states}
	
\author{Junpeng Liu}
\affiliation{ School of Physical Science and Technology, Lanzhou University, Lanzhou 730000, China}
\affiliation{ Lanzhou Center for Theoretical Physics and Key Laboratory of Theoretical Physics of Gansu Province, Lanzhou University, Lanzhou 730000, China}
\author{Miaomiao Zhao}
\affiliation{ School of Physical Science and Technology, Lanzhou University, Lanzhou 730000, China}
\affiliation{ Lanzhou Center for Theoretical Physics and Key Laboratory of Theoretical Physics of Gansu Province, Lanzhou University, Lanzhou 730000, China}
\author{Yun-Tong Yang}
\affiliation{ School of Physical Science and Technology, Lanzhou University, Lanzhou 730000, China}
\affiliation{ Lanzhou Center for Theoretical Physics and Key Laboratory of Theoretical Physics of Gansu Province, Lanzhou University, Lanzhou 730000, China}
\author{Hong-Gang Luo}
\email{luohg@lzu.edu.cn}
\affiliation{ School of Physical Science and Technology, Lanzhou University, Lanzhou 730000, China}
\affiliation{ Lanzhou Center for Theoretical Physics and Key Laboratory of Theoretical Physics of Gansu Province, Lanzhou University, Lanzhou 730000, China}
\affiliation{ Beijing Computational Science Research Center, Beijing 100084, China}

\date{\today }

\begin{abstract}		
The ground state of quantum Rabi model (QRM) exhibits rich nonclassical states including squeezed state, cat state, and entangled state in different parameter regimes. In this paper, we firstly use the polaron picture to figure out the process of superradiant phase transition  (SPT) in view of the nonclassical states.
According to the polaron wave functions, the squeezed state extends to whole parameter regimes, and a small but non-zero weighted antipolaron is necessary to form novel semi-cat states.
Moreover, the behavior of superradiance in the QRM can be viewed as the same displacement of the cat states from zero to a finite value, while the ground state becomes entangled state resulting from the emergence of spin-up state. On the other hand, the nonclassical states can be intuitively characterized by the Wigner functions with analytical expressions in the polaron picture, and the Wigner negativity is also evaluated to measure the nonclassicality.
Based on the squeezing and superradiance, a classification of the coupling strength for the nonclassical states containing in the ground state is presented, and the process of the SPT is also revealed by the photon number distribution in Fock space, which is important to understand the fundamental quantum physics in the QRM. Our work provides a guidance for preparing the nonclassical states in experiments based on the QRM.
\end{abstract}

\maketitle

\preprint{APS/123-QED}

\section{\label{sec:level1}Introduction}
The quantum Rabi model  (QRM) \cite{Rabi1937PR} plays an important role in understanding the fundamental physics in the field of light-matter interaction, which describes the coupling of a harmonic oscillator with frequecy $\hbar\omega$ for a single-mode light field in the cavity and a two-level atom with frequency $\Delta$ and the coupling strength $g$.
Despite its simple form, the analytical solutions with closed-form expression for
the general energy eigenvalues of the QRM are still difficult to obtain \cite{Forn2019RMP,Li2021PRA}, which hinders physical exploration of the QRM.
The well-known Jaynes-Cummings  model (JCM) \cite{Jaynes1963PIEEE} is obtained by making rotating-wave approximation(RWA), but is only justified for near-resonant ($\hbar\omega \approx \Delta$) and weakly coupled ($g \ll \hbar\omega$) parametric regions.
For the cavity quantum electrodynamics (QED), the coupling strength is usually limited, and in this case the JCM is successful in understanding a range of experimental phenomena, such as quantum Rabi oscillation \cite{Brune1996PRL} and vacuum Rabi mode splitting \cite{Thompson1992PRL}.
Thanks to flexible tunability of various artificial systems, quantum simulation can lead to further increase in coupling strength, and then RWA is invalid when the coupling strength became comparable to or is even greater than the transition frequencies in the system, corresponding to ultra-strong coupling ($g \gtrsim 0.1\hbar\omega$) \cite{Anappara2009PRB,Niemczyk2010NP, FornDiaz2017NP} and deep-strong coupling ($g \gtrsim \hbar\omega$) \cite{Liberato2014PRL, Langford2017NC, Yoshihara2017NP}, respectively.
One notes that this classification is just a historical convention, and there is no any deeper physical meaning \cite{Frisk2019NRP}.
While the classification based on the near-resonant condition is widely applicable \cite{Rossatto2017PRA}, it is not suitable for other parametric regions, such as superradiant phase transition (SPT) with the condition of  ratio of the frequencies $R\equiv \Delta/(\hbar\omega) \rightarrow \infty$ \cite{Hwang2015PRL,Liu2017PRL}.

Recently, the SPT  has been experimentally simulated by  a single trapped ion \cite{Cai2021NC,Cai2022CPL} and  a nuclear magnetic resonance \cite{Chen2021NC}, respectively, with large ratios $R=100$ and $R=50$.
Beside the SPT occurring in ground state of the QRM as the coupling strength increases across the critical coupling strength, there are several kinds of nonclassical states existing in corresponding parametric regions, including squeezed state,  cat state and entangled state \cite{Ashhab2010PRA, Leroux2017PRA, Chen2020PRA}.
The nonclassical states have wide applications in modern quantum technologies, such as the cat states for quantum computing \cite{Hacker2019NP,Bergmann2016PRA,Grimm2020Nature}, the entangled states for quantum information processing \cite{Li2017PRL,Albert2016PRL,Sun2021PRL} and the squeezed states for quantum metrology \cite{Joo2011PRL,Hastrup2021PRL,Sanchez2021PRL,Xin2021PRL}.
In this paper, we explore a classification of parametric regions suitable for the SPT with well-defined physical meanings in view of the properties of the nonclassical states, and figure out  corresponding process of the SPT by characterizing the photon populations in Fock space \cite{Yang2022CharacterizingSP}, which are important for understanding the fundamental quantum physics and provide a guidance for preparing the nonclassical states in experements based on the QRM.

The polaron picture is used to investigate the properties of the nonclassical states contained in the ground state for the superradiant regime  of the QRM.
It is a useful method for extracting the ground-state wave function of the QRM \cite{Ying2015PRA, Cong2017PRA} and its extensions \cite{Cong2019PRA, Sun2020PRA, Ying2022QUTE}. Here we improve the method by reducing the variational parameters, and use it to analyze the properties of the nonclassical states in all parametric ranges, then the squeezed region is distinguished and the exotic semi-cat states are found.
In addition, the Wigner function with analytical expressions is derived to characterize the nonclassical states, and  the Wigner negativity is also evaluated to measure the nonclassicality.
In this perspective, we obtain a classification of the coupling strength for the nonclassical states and reveal the process of SPT: the transition between different nonclassical states.

The paper is organized as follows. In Sec. \ref{sec:level2}, the improved polaron picture with high precision is  given and the analytical expressions of the Wigner function are derived to characterize the nonclassical states, we find that the squeezing occurs in a specific parameter region and the antipolaron is necessary for leading to the semi-cat states.
In Sec. \ref{sec:level3},  the $x$-type SPT is dominated the cat states with the same displacement,  the  Wigner negativity and the entanglement entropy  are calculated to measure the  nonclassicality and the entanglement for the ground state, respectively.
In Sec. \ref{sec:level4}, we discuss the nonclassical states corresponding to different coupling strength intervals based on the squeezing and superradiance, and obtain a classification to understand the process of the SPT by characterizing the photon populations in Fock space.
A brief conclusion is finally given in Sec. \ref{Conclusion}.

\section{\label{sec:level2}The Polaron Picture for the QRM}
The QRM describes a quantized harmonic oscillator for a single-mode light field in the cavity coupled to a quantum two-level system for the atom, the Hamiltonian reads
\begin{equation}\label{H}
	\hat{H}= \hbar \omega \hat{a}^{\dagger }\hat{a}+\frac{\Delta }{2}\hat{\sigma} _{z}+g\hat{\sigma} _{x}(\hat{a}^{\dagger
	}+\hat{a}),
\end{equation}
where $\hat{a}^{\dagger }$ and $\hat{a}$ are respectively the creation and annihilation operator  for the harmonic oscillator with
frequency $\hbar\omega $, the two-level system is described by Pauli matrices $
\hat{\sigma}_{x,z}$ with the frequency $\Delta $,  $g$ is coupling strength between the two-level system and the quantum oscillator. And the dimensionless ratio parameter is defined as $R\equiv\Delta/(\hbar\omega)$.
For the SPT, the critical point $g_c=\sqrt{(\hbar\omega)^{2}+\sqrt{(\hbar\omega)^{4}+g_{c0}^{4}}}$ is associated with the  maximum of squeezing in the polaron picture as an improved  coupling scale compared to  $g_{c0}=\sqrt{\hbar\omega\Delta}/2$, which is  applied to a wider range of parameters \cite{Ying2015PRA}.

\subsection{\label{sec:level2A} The wave function and energy of the ground state}
In terms of the creation and annihilation operators of quantum harmonic oscillator with dimensionless formalism $\hat{a}^{\dagger}=(\hat{x}-i\hat{p})/\sqrt{2}$, $\hat{a}=(\hat{x}+i\hat{p})/\sqrt{2}$, where $\hat{x}=x$ and $\hat{p}=-i\frac{\partial}{\partial x}$ denote the position and momentum operators, respectively, the Hamiltonian (\ref{H}) becomes
\begin{equation}\label{HP}
\hat{H}^\prime=\sum\limits_{\sigma _{x}=\pm }\left( \hat{h}^{\sigma _{x}}\left\vert \sigma
_{x}\right\rangle \left\langle \sigma _{x}\right\vert +\frac{\Delta }{2}%
\left\vert \sigma _{x}\right\rangle \left\langle \overline{\sigma }%
_{x}\right\vert \right) +\varepsilon _{0},
\end{equation}
where $\overline{\sigma}_{x}=-\sigma_{x}$, and $+$($-$) labels the state $\left|\uparrow\right\rangle_{x}$ ($\left|\downarrow\right\rangle_{x}$) of spin in the $\sigma_{x}$ representation.
$\hat{h}^{\pm}=\hbar \omega (\hat{p}^{2}+\hat{v}_{\pm})/2$, where $\hat{v}_{\pm}=(\hat{x} \pm g^{\prime2})$, $g^{\prime}=\sqrt{2}g / (\hbar\omega)$, and $\varepsilon _{0}=- \hbar \omega (g^{\prime 2}+1)/2$.

The ground-state wave function $\Psi_{x}$ satisfies the Schr\"{o}dinger equation $H \Psi_{x}=E\Psi_{x} $,
then it should take the form of
\begin{equation}\label{WFx}
	\Psi_{x}=\frac{1}{\sqrt{2}}(\psi_{+}\left|\uparrow\right\rangle_{x} - \psi_{-}\left|\downarrow\right\rangle_{x}),
\end{equation}
where $\psi_{\pm}=\psi(\pm x)$.
The trial variational wave function for $\psi (x)$
takes the superposition of the deformed polaron $\varphi_{\alpha}$ and
antipolaron $\varphi_{\beta}$,
\begin{equation}\label{wavefunction}
	\begin{aligned}	
	\psi \left(x\right)&=\alpha \varphi_{\alpha}(x)+\beta \varphi_{\beta}(x)
	\\	&=\alpha \left(\frac{\xi}{\pi}\right)^{\frac{1}{4}}\exp\left[-\frac{\xi\left(x-D_{\alpha}\right)^{2}}{2} \right]
\\ &+\beta \left(\frac{\xi}{\pi}\right)^{\frac{1}{4}}\exp\left[-\frac{\xi\left(x+D_{\beta}\right)^{2}}{2} \right],
	\end{aligned}	
\end{equation}
where $D_{i}=\zeta_{i}g^{\prime}$ denote displacements, and $i=\alpha,\beta$ are respectively weights of the polaron and antipolaron with the same squeezed parameter $\xi$, which differs from two polarons with different squeezed parameters in Ref. \cite{Ying2015PRA}.
Note that $\psi_{\pm}$ in the ground-state wave function (\ref{WFx}) are  generalized cats when the polarons have nonzero displacements, which allows different weights for the coherent contributions, and satisfies that $\psi_{+}(x)=\psi_{-}(-x)$.

\begin{figure}[htbp]
	\centering
	\includegraphics[width=1.05\linewidth]{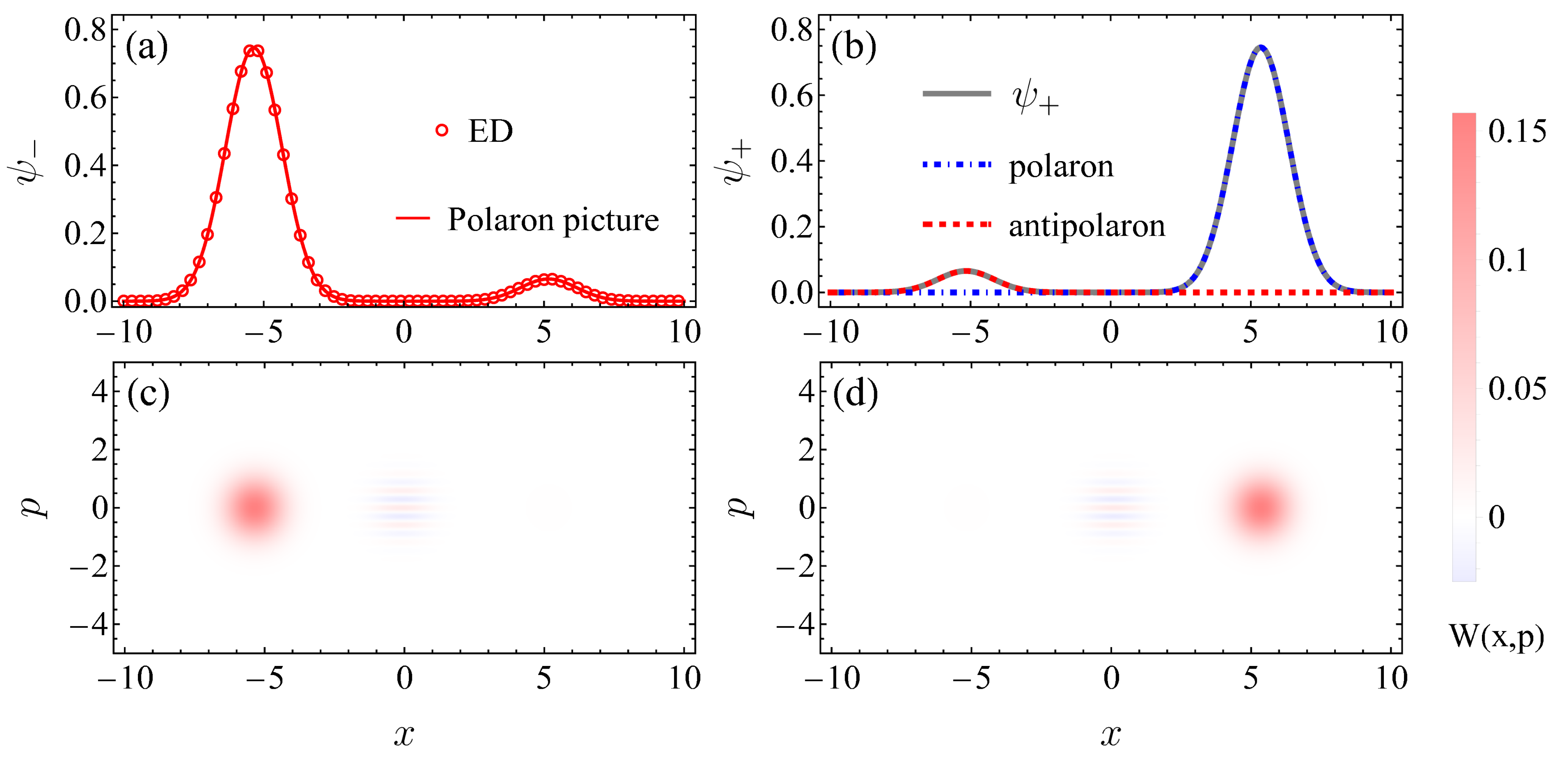}
	\caption{The wave functions and  corresponding  Wigner functions in $\sigma_{x}$ representation by using polaron picture for ground state of the QRM.
		(a) The wave function of the state $\psi_{-}$, which is in agreement with ED; and (b) the wave function of the state $\psi_{+}$ consists of polaron and antipolaron.
		(c) Wigner function $W_{x}^{-}$ of the state $\psi_{-}$,  and (d) Wigner function $W_{x}^{+}$ of the state $\psi_{+}$.
		The parameters are $R=10$, $g/g_{c}=2.0$, $\alpha=0.996$, $\beta=0.087$.}
	\label{wfWFx}
\end{figure}%

\begin{figure}[htbp]
	\centering
	\includegraphics[width=0.8\linewidth]{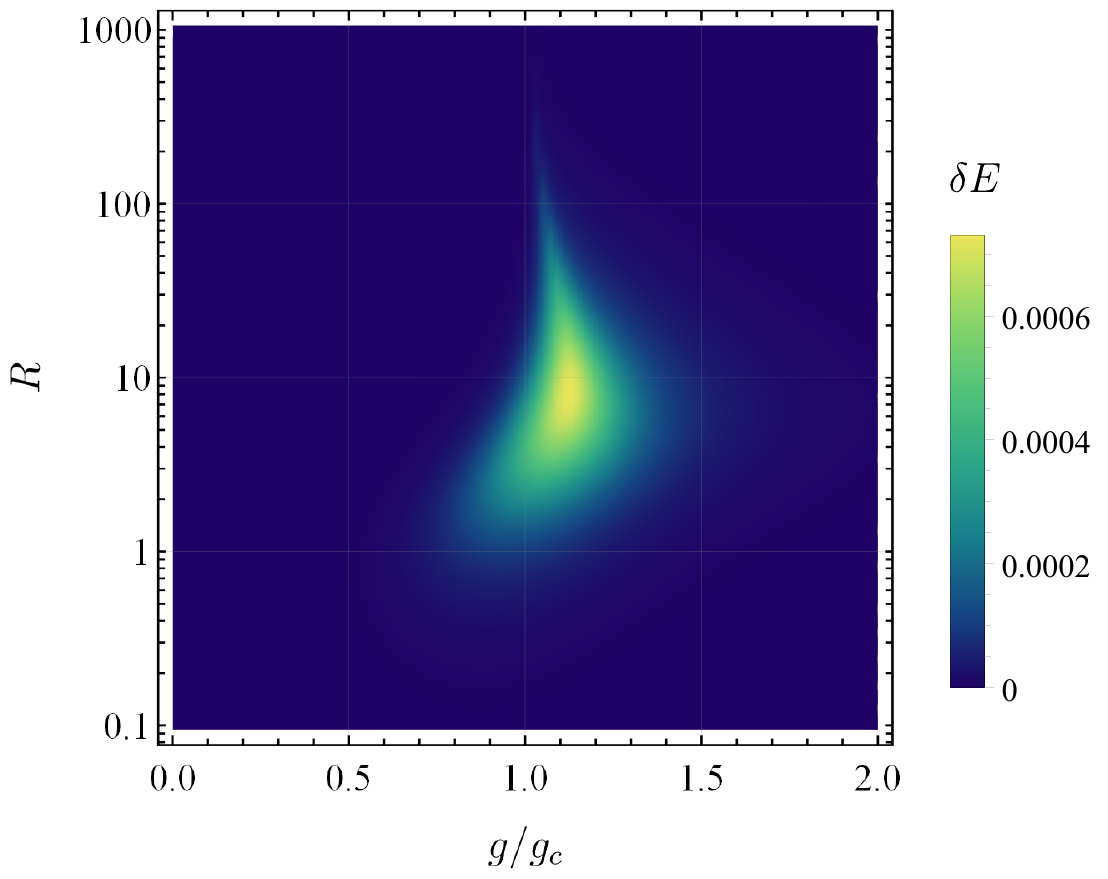}
	\caption{The energy error as functions of the coupling strength $g/g_{c}$  and the ratio parameter $R$ for ground state of the QRM in the polaron picture we considered in the present work. In the most parameter regime the polaron picture works well.}
	\label{Eerror}
\end{figure}

The wave function Eq. (\ref{WFx}) satisfies the normalization condition, i.e.
\begin{equation}\label{nc}
	\langle \Psi_{x} |\Psi_{x}\rangle=\langle \psi_{\pm} |\psi_{\pm} \rangle=1,
\end{equation}
thus the probabilities of the two-level system in state $\left|\uparrow\right\rangle_{x}$ and $\left|\downarrow\right\rangle_{x}$ are independent on the parameters.
The weight of the antipolaron  $\beta$ can be solved by the normalization condition Eq. (\ref{nc}), then only $4$ independent variational parameters $\{\alpha,\xi,\zeta_{\alpha},\zeta_{\beta}\}$
are required by extracting from the energy minimization
\begin{equation}\label{vm}
	\frac{\partial E}{\partial \alpha}=\frac{\partial E}{\partial \xi}=\frac{\partial E}{\partial \zeta_{\alpha}}=\frac{\partial E}{\partial \zeta_{\beta}}=0,
\end{equation}
and
\begin{widetext}
\begin{equation}\label{E}
		\begin{aligned}
		E=&\langle \Psi |H|\Psi\rangle=\langle \psi_{+} |h^{+}|\psi_{+}\rangle - \frac{ \Delta}{2} \langle \psi_{+}|\psi_{-} \rangle+\varepsilon_{0} \\
		=&\frac{\hbar\omega}{2}[\alpha^{2} (U^{\alpha}_{-\alpha}+V^{\alpha}_{-\alpha})+\beta^{2} (U^{-\beta}_{\beta}+ V^{\beta}_{-\beta})
         +2\alpha\beta( U^{\alpha}_{\beta} + V^{\alpha}_{\beta})]-\frac{\Delta}{2}(\alpha^{2} +\beta^{2} +2\alpha\beta  T^{\alpha}_{\beta} )+\varepsilon_{0},
	\end{aligned}
\end{equation}
\end{widetext}
where $F^{\pm i}_{\pm j}=F(\pm D_{i},\pm D_{j})$,
is given by the functions
\begin{equation}
	\begin{aligned}
		T(D_{i},D_{j})&=\exp[-\frac{\left(D_{i}+D_{j}\right)^{2}\xi}{4}],\\
		U(D_{i},D_{j})&=T(D_{i},D_{j})\frac{2+\xi(2 g^{\prime2}-D_{i}+D_{j})^{2}}{4\xi},\\
		V(D_{i},D_{j})&=T(D_{i},D_{j})\frac{2\xi-\xi^{2}(D_{i}+D_{j})^{2}}{4}.
	\end{aligned}
\end{equation}
Fig. \ref{wfWFx}(a) and (b) show the wave functions for the states $\psi_{\mp}$, respectively.
The energy error between the polaron picture and exact result
\begin{equation}
\delta E= \left| \frac{E-E_{exact}}{E_{exact}} \right|,
\end{equation}
is obtained within all parameter ranges as shown in Fig. \ref{Eerror}, which has an accuracy up to $10^{-4}$.
The higher accuracy can be obtained by using the two-polaron picture with $5$ independent variational parameters \cite{Ying2015PRA} or even multipolaron expansion method with more variational parameters \cite{Cong2017PRA}, however, the two-polaron picture with $4$ independent variational parameters in this paper indicates that the state optimally captures the essential physics necessary to accurately describe the ground state of the QRM.

\begin{figure}[htbp!]
	\centering
	\includegraphics[width=0.78\linewidth]{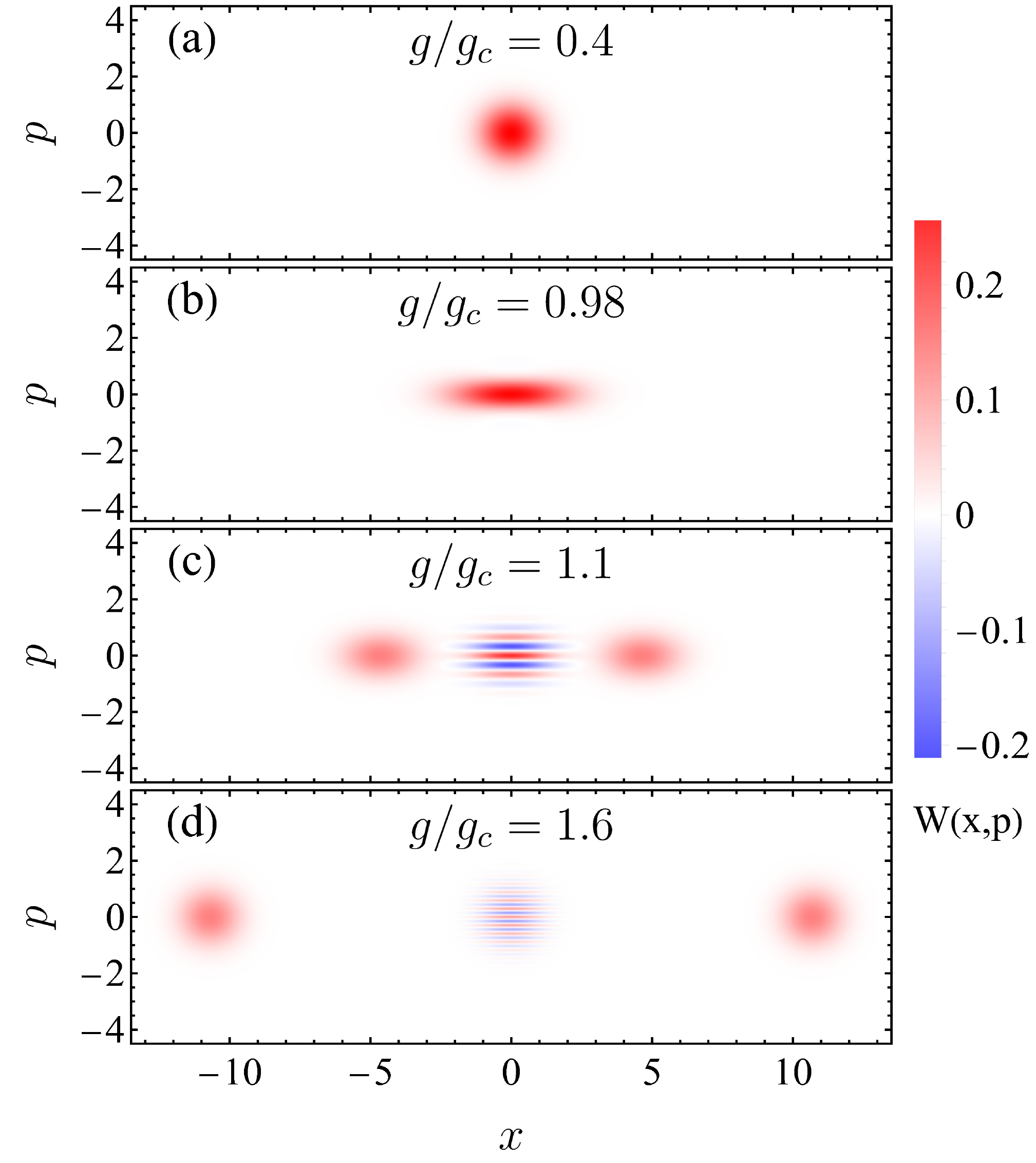}
	\caption{The Wigner function $W(x,p)$ for the ground state by using the polaron picture for different coupling strengths.
		It can be seen that these characteristics are provided with the nonclassical states: (a) vacuum states, (b) squeezed vacuum states, (c) squeezed cat states, and (d) cat states without squeezing.
		The parameters are $R=100$, (a) $g/g_{c}=0.4$, (b) $g/g_{c}=0.98$, (c) $g/g_{c}=1.1$, and (d) $g/g_{c}=1.6$.}
	\label{WFt}
\end{figure}

To characterize the ground state of the QRM, we calculate the Wigner function
\begin{equation}
	W(x,p)=\frac{1}{\pi} \int_{-\infty}^{\infty} \mathrm{d}x^\prime \psi(x-x^{\prime})\psi(x+x^{\prime})e^{2ix^{\prime}p}.
\end{equation}
And the expressions of the Wigner function for the wave functions $\psi_{\pm}$  in $\sigma_{x}$ representation can be obtained as
\begin{equation}\label{WignerFx}
	\begin{aligned}
		W^{+}_{x}=&W_{\alpha}^{R}+W_{I}^{+}+W_{\beta}^{L},\\
		W^{-}_{x}=&W_{\alpha}^{L}+W_{I}^{-}+W_{\beta}^{R},\\
		\end{aligned}
\end{equation}
where 		
\begin{equation}
	\begin{aligned}
		W_{\alpha}^{R}=&\frac{1}{2\pi}\alpha^{2}N^{-\alpha}_{-\alpha},\quad W_{\beta}^{L}=\frac{1}{2\pi}\beta^{2}N^{\beta}_{\beta},\\
		W_{\alpha}^{L}=&\frac{1}{2\pi}\alpha^{2}N^{\alpha}_{\alpha}, \quad W_{\beta}^{R}=\frac{1}{2\pi}\beta^{2}N^{-\beta}_{-\beta},\\
		W_{I}^{+}=&\frac{1}{\pi}\alpha\beta N^{\alpha}_{-\beta} M^{\alpha}_{\beta}, \quad W_{I}^{-}=\frac{1}{\pi}\alpha\beta N^{-\alpha}_{\beta} M^{-\alpha}_{-\beta},
	\end{aligned}
\end{equation}
and $F^{\pm i}_{\pm j}=F(\pm D_{i},\pm D_{j})$ is given by the functions
\begin{equation}
	\label{NM}
	\begin{aligned}	
		N(D_{i},D_{j})&=\exp\left[-\frac{(2p)^{2}+\left(2x+D_{i}+D_{j}\right)^{2}\xi^{2}}{4\xi}\right],\\
		M(D_{i},D_{j})&=\cos\left[\left(D_{i}+D_{j}\right)p\right].
	\end{aligned}
\end{equation}
Note that $N(D_{i},D_{j})$  determine  areas of the Wigner functions, $W_{\alpha}^{R}$ ($W_{\alpha}^{L}$) and $W_{\beta}^{R}$   ($W_{\beta}^{L}$) are  produced by the polaron $\varphi_{\alpha}$ and the antipolaron $\varphi_{\beta}$, respectively, and are on the right (left) side of the Wigner function, often referred to as states of ``alive cat (dead cat)''.
$W_{I}^{\pm}$ are the interference terms, where $M(D_{i},D_{j})$  determines  oscillations of the Wigner functions.
And intensities of the Wigner functions depend on the weights $\alpha$ and $\beta$. Fig. \ref{wfWFx} (a,b) and (c,d) show the wave functions and corresponding Wigner functions for the states $\psi_{\mp}$, respectively.
It turns out to be that the Wigner functions satisfy $W^{\pm}_{x}(x,p)=W^{\mp}_{x}(-x,p)$, since the wave functions  satisfy $\psi_{+}(x)=\psi_{-}(-x)$.
Therefore, the total Wigner function of the ground state  $W_{T}=W_{x}^{+}+W_{x}^{-}$, as shown in Fig. \ref{WFt}.

\subsection{\label{sec:level2B} Squeezed region: Maximum squeezing near the critical point for the significant ratio $R$}
From Eq. (\ref{NM}), the expression of the area term of Wigner functions  depends on the squeezing parameters $\xi$, the term containing coordinates grows more slowly than the term containing momentum if $\xi<1$,  thus causing the squeezing in the $p$ direction (see Fig. \ref{WFt}(b,c)), and the smaller $\xi$, the larger the squeezing.

\begin{figure}[htp]
	\centering
	\includegraphics[width=0.85\linewidth]{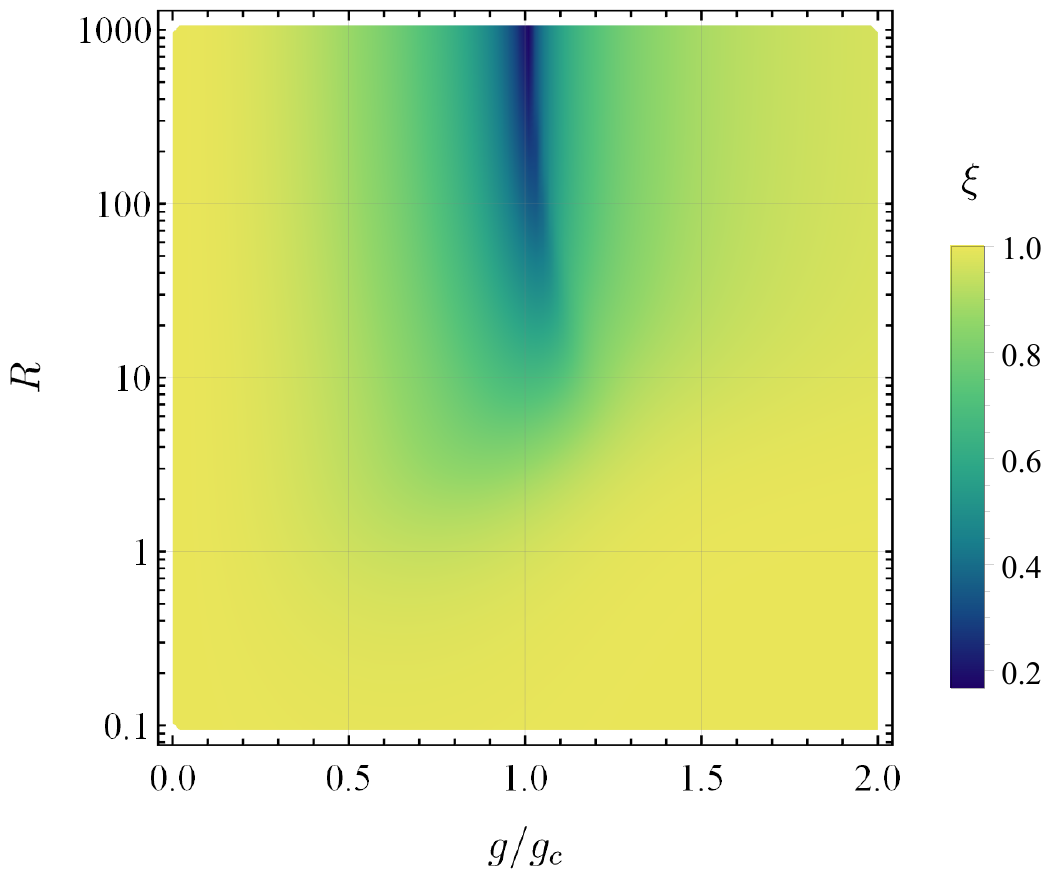}
	\caption{The squeezing parameter $\xi$ as a function of the coupling strength $g/g_{c}$ and the ratio parameter $R$.}
	\label{RGS}
\end{figure}

Fig. \ref{RGS} shows the squeezing parameter $\xi$ as a function of the coupling strength $g/g_{c}$ and the ratio parameter $R$ in all parameter ranges.
When the ratio parameter $R$ is small,  there is almost no squeezing in all regimes of the coupling strength; when the ratio parameter $R$ is large,  the squeezing occurs in the coupling strength interval of $0.5\lesssim g/g_{c}\lesssim1.5$, and reaches a maximum near the critical point $g_{c}$.
Around the critical point, the larger the ratio parameter $R$, the greater the squeezing.
To sum up, for the significant $R$, with the increase of coupling strength, the squeezing increases continuously before the critical point, and reaches the maximum value near the critical point,  then decreases  after the critical point, finally vanishes when the coupling strength becomes strong enough.

It is worth mentioning that the maximum squeeze at the critical point is crucial for the development of criticality-enhanced quantum metrology \cite{Chu2021PRL,Ying2022Entropy}.

\subsection{\label{sec:level2C}The necessary antipolaron with very small but nonzero weight leading to semi-cat states}
\begin{figure}[htbp!]
	\centering
	\includegraphics[width=0.78\linewidth]{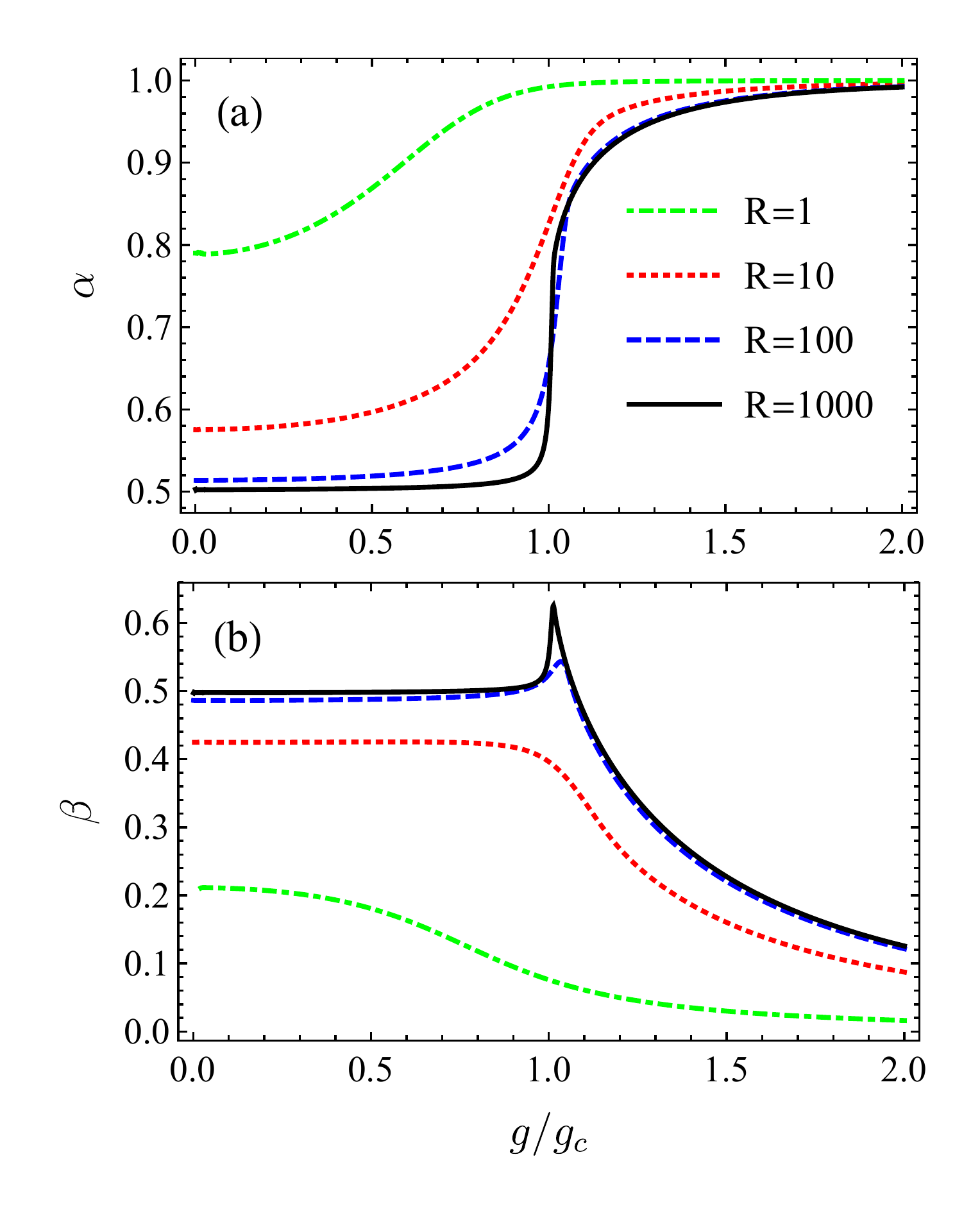}
	\caption{The weights of the polaron (a) and the antipolaron (b) as functions of the coupling strength $g/g_{c}$ for different $R$. The green dotted-dash, red dot, blue dash and black solid lines denote the weights for the ratio parameters $R=1$, $10$, $100$ and $1000$, respectively. }
	\label{Weight}
\end{figure}

In $\sigma_{x}$ representation, $\psi_{\pm}$ in the ground-state wave function (\ref{WFx}) are  generalized cats \cite{Hacker2019NP} when the polarons have nonzero displacements, which allow different weights for the coherent contributions.
As shown in Fig. \ref{Weight}, (a) the weight of the polaron $\alpha \rightarrow 1$, and (b) the weight of the  antipolaron $\beta$ becomes small but nonzero when the coupling strength $g/g_{c} \gtrsim 1.5$.
According to Eq. (\ref{WignerFx}), the intensities of Wigner functions depend on the weights, and the intensities of interference term $W_{I}^{\pm}$ depends on product of the weights $\alpha\beta \rightarrow \beta$, meanwhile, $\beta^{2} \rightarrow 0$,
so that the Wigner functions  belonging to the antipolaron in $W^{\pm}_{x}$ vanish, i.e.
\begin{equation}
	\begin{aligned}
		W^{+}_{x}\rightarrow W^{Alive}_{x} \equiv  &W_{\alpha}^{R}+W_{I}^{+},\\
		W^{-}_{x}\rightarrow W^{Dead}_{x} \equiv   &W_{\alpha}^{L}+W_{I}^{-},\\
	\end{aligned}
\end{equation}
where 	$W^{Alive}_{x}$ and	$W^{Dead}_{x}$ denote  alive-semi-cat state  and  dead-semi-cat state, which are the even cat states that exclude only ``dead cat'' part and ``alive cat'' part as shown in  Fig. \ref{wfWFx}(c,d), respectively.
It can be seen that the interference fringes are clearly visible, which demonstrates that the composition of the small but nonzero antipolaron in the wave function (\ref{wavefunction}) is necessary,  when the coupling strength is strong enough.

In particular, as shown by the green dotdashed line in  Fig. \ref{Weight}(b), for the case of smaller ratio parameter, the weight $\beta \rightarrow 0$  under the enough strong coupling strength, leading to $W_{I}^{\pm} \rightarrow 0$, the interference fringes fade away, and the semi-cat states turn into corresponding coherent states,  the ground-state wave function (\ref{WFx}) becomes
\begin{equation}\label{WFx1}
	\Psi_{x}=\frac{1}{\sqrt{2}}\left[\varphi_{\alpha}(x)\left|\uparrow\right\rangle_{x} - \varphi_{\alpha}(-x)\left|\downarrow\right\rangle_{x}\right],
\end{equation}
where only one variational parameter $\{\zeta_{\alpha}\}$ is required.

\section{\label{sec:level3} The nonclassical states within the SPT}
The nonclassical states contained in the ground state of the QRM include squeezed state, cat state, and entangled state.
Fig. \ref{WFt} shows the total Wigner functions for different coupling strength  with the ratio parameter $R=100$, it can be seen that the ground state exhibits the properties of (a) vacuum state, (b) squeezed vacuum state,  (c) squeezed cat state, and (d) cat state without squeezing, as the coupling strength increases.
In the following, we will analyze the variational parameters and calculate the related physical quantities to demonstrate the properties of these states as functions of the coupling strength and the ratio parameter.

\begin{figure}[hbtp!]
	\centering
	\includegraphics[width=1.05\linewidth]{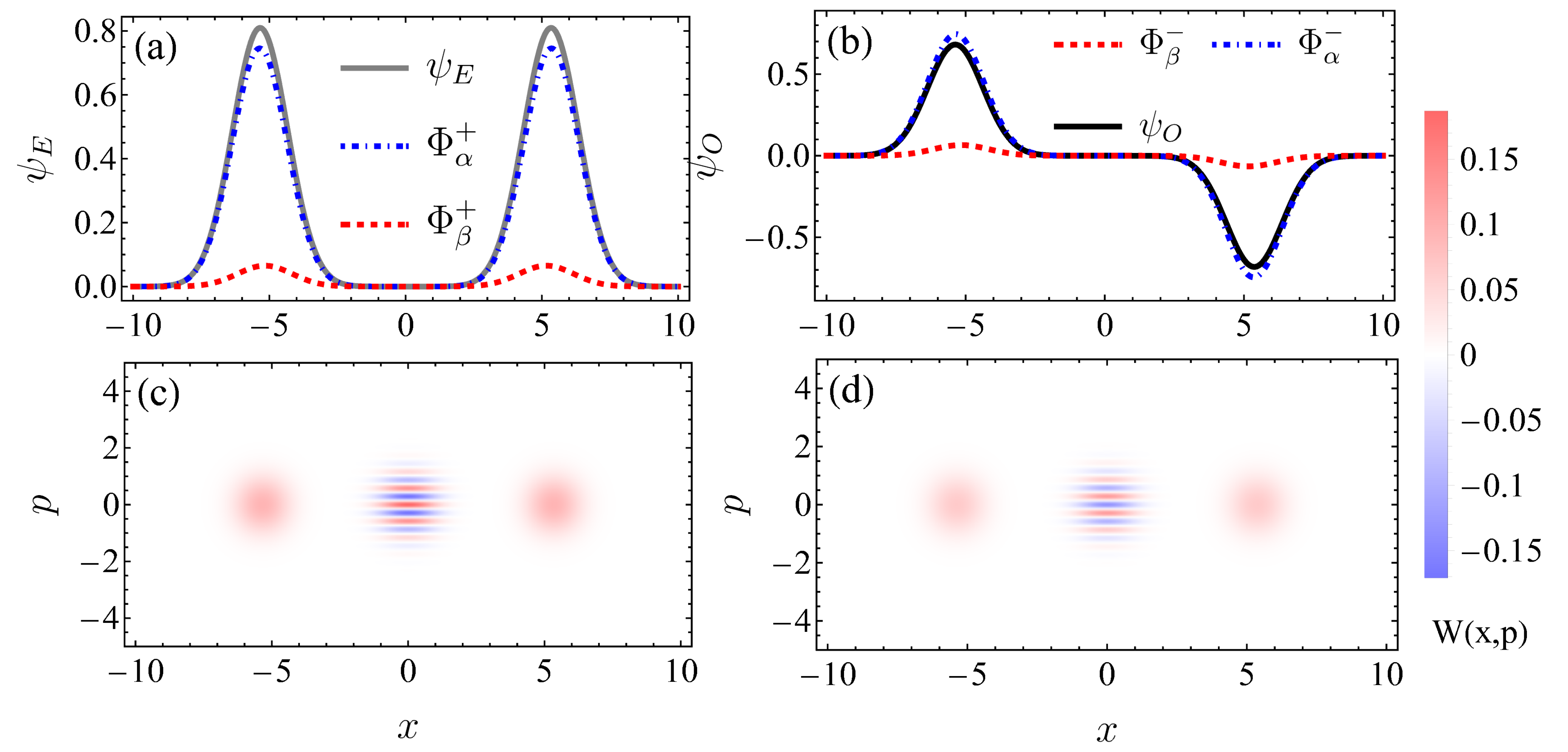}
	\caption{The wave functions and  corresponding  Wigner functions in $\sigma_{z}$ representation by using polaron picture for ground state of the QRM.
		(a) The wave functions of the even cat states,  and (b) the wave functions of the odd cat states in the Eq. (\ref{WFz});
		(c) the Wigner function $W_{z}^{E}$ for the even cat state $\psi_{E}$ entangled with $\left|\downarrow\right\rangle_{z}$ and (d) $W_{z}^{O}$ for the odd cat state $\psi_{O}$ entangled with $\left|\uparrow\right\rangle_{z}$.
		The parameters are the same as in Fig. \ref{wfWFx}.}
	\label{WFZ}
\end{figure}

\subsection{\label{sec:level3A} The $x$-type SPT caused by the cavity cat states}
In $\sigma_{z}$ representation, through  Schmidt decomposition  the form of ground-state wave function (\ref{WFx}) of the QRM becomes
\begin{equation}\label{WFz}
	\Psi_{z}=\frac{1}{2}\left(\psi_{O}\left|\uparrow\right\rangle_{z}+\psi_{E}\left|\downarrow\right\rangle_{z} \right),
\end{equation}
where
\begin{equation}
	\begin{aligned}
	\psi_{O}=&\psi_{+} - \psi_{-}=\alpha\Phi^{-}_{\alpha}-\beta\Phi^{-}_{\beta}, \\
	\psi_{E}=&\psi_{+} + \psi_{-}=\alpha\Phi^{+}_{\alpha}+\beta\Phi^{+}_{\beta},
	\end{aligned}
\end{equation}
and $\Phi^{\pm}_{i}=\varphi_{i}(x) \pm \varphi_{i}(-x)$ respectively denote even cat state (containing only even photon numbers) and odd cat state (containing only odd photon numbers) if $D_{i}> 0$.
$\psi_{O}$ and $\psi_{E}$ are the linear superposition of the odd cat states and the even cat states (see Fig. \ref{WFZ}(a,b)), respectively,  entangled with the spin-up state $\left|\uparrow\right\rangle_{z}$ and -down state $\left|\downarrow\right\rangle_{z}$ in Eq. (\ref{WFz}), corresponding to the ground state  belonging to  the even parity chain \cite{Casanova2010PRL}:
\begin{equation}
	\left|\downarrow,0 \right\rangle_{z}\leftrightarrow\left|\uparrow,1 \right\rangle_{z}\leftrightarrow\left|\downarrow,2 \right\rangle_{z}\leftrightarrow\left|\uparrow,3 \right\rangle_{z}\leftrightarrow \cdots(p=+1).
\end{equation}

The Wigner functions for the cat states $\psi_{O}$ and $\psi_{E}$ can be obtained by
\begin{equation}\label{WOE0}
	\begin{aligned}
		W^{O}_{z}=&\frac{1}{2}W_{T}-W_{D},\\
		W^{E}_{z}=&\frac{1}{2}W_{T}+W_{D},
	\end{aligned}
\end{equation}
where
\begin{equation}
	\begin{aligned}
		W_{D}=&	\frac{1}{2\pi}[\alpha^{2}N^{0}_{0}M_{\alpha}^{\alpha} +\beta^{2}N^{0}_{0}M_{\beta}^{\beta} \\
		&+\alpha\beta(N^{-\alpha}_{-\beta} M^{-\alpha}_{\beta}+N^{\alpha}_{\beta}M^{\alpha}_{-\beta})].
	\end{aligned}
\end{equation}
Fig. \ref{WFZ} (a,b) and  (c,d) show the wave functions and corresponding Wigner functions for the two kinds of  cat states, respectively.
Moreover,  the weight of the odd cat states always is less than that of the even cat states  in superradiant phase due to the nonzero $\beta$ (see Fig. \ref{Weight} (b)), according to Eq. (\ref{WFz}), the total Wigner function of the ground state indicates the characteristic of the even cat states as shown in Fig. \ref{WFt}.

\begin{figure*}[htbp!]
	\centering
	\includegraphics[width=0.9\linewidth]{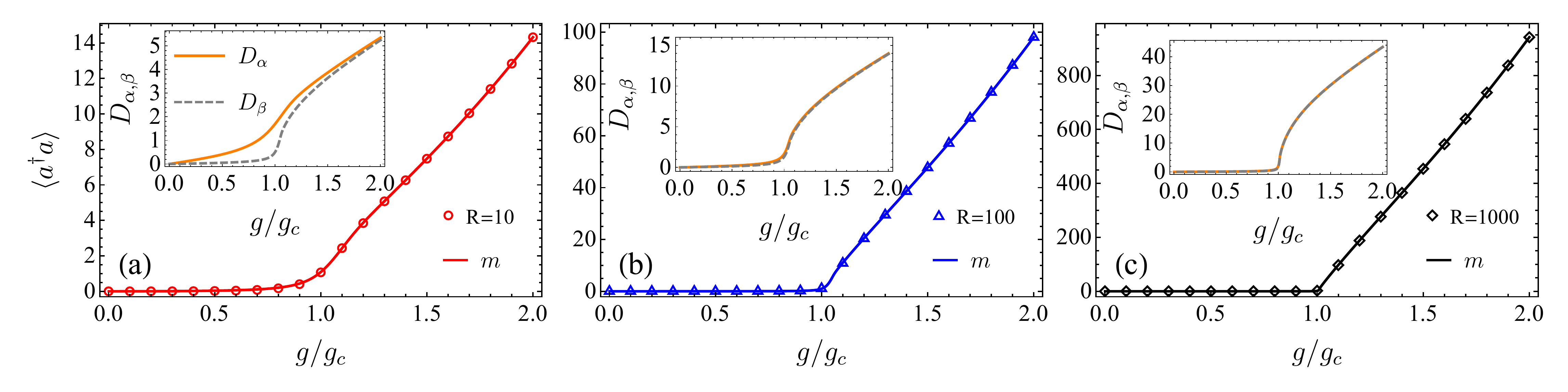}
	\caption{ (a-c) The exact mean photon number (symbols)  versus  $m$ (solid lines) as functions of the coupling strength $g/g_{c}$ of the ground state  for  $R=10$, $R=100$ and $R=1000$, respectively. The symbols denote the numerically exact results. The insets are  the displacements using the polaron picture as functions of the coupling strength $g/g_{c}$ for the different ratio parameter $R$, the orange solid and gray dashed lines denote the displacements  $D_{\alpha}$ and $D_{\beta}$, respectively.}
	\label{DMPN}
\end{figure*}

The mean photon number of the ground state are always used as the order parameters of the SPT in the QRM \cite{Hwang2015PRL,Liu2017PRL}, which undergo the abrupt changes from $0$ to finite numbers when the coupling strength goes through the critical point $g_{c}$ in the limit $R\rightarrow\infty$ (see Fig. \ref{DMPN}), and the phenomena have been observed experimentally \cite{Cai2021NC,Chen2021NC}.
Note that the $x$-type SPT for the ground state of the QRM has been found in  Ref. \cite{Liu2017PRL}, and the mean photon number of  Eq. (\ref{WFz}) can be given by $m=(\alpha^{2}D_{\alpha}^{2}+\beta^{2}D_{\beta}^{2})/2$.
In Fig. \ref{DMPN}, comparing the exact mean photon number (symbols)  and  $m$ (solid lines), it can be seen that they are very consistent  for different ratio parameter $R$.
As shown in the insets of Fig. \ref{DMPN}, when the system from the normal phase enters to the superradiant phase, the displacements of two polarons undergo abrupt changes from $0$ to finite values  for significant ratio parameter $R$.
In the normal phase, the displacements $D_{\alpha,\beta} \rightarrow 0$, which leads to  the vacuum states for the quantum oscillator  (see Fig. \ref{WFt} (a,b)), and the nonzero displacements lead to the cat states in the superradiant phase (see Fig. \ref{WFt} (c,d)).
From Fig. \ref{RGS}, the ground state is squeezed around the critical point, thus there are the squeezed vacuum state before the critical point and the squeezed cat states after the critical point as shown in Fig. \ref{WFt} (b,c), respectively.
And for the large $R$, the displacement difference between the two polarons  $\delta D=\left| D_{\alpha} - D_{\beta}\right| \rightarrow 0$, thus the behavior of superradiance  in the QRM can be viewed as being caused by the same displacement of the cat states from zero to a finite value, i.e.
\begin{equation}
	\label{awf}	\Psi_{z}=\frac{1}{2}\left[\left(\alpha-\beta\right)\Phi^{-}_{\alpha}\left|\uparrow\right\rangle_{z} +\left(\alpha+\beta\right)\Phi^{+}_{\alpha}\left|\downarrow\right\rangle_{z} \right],
\end{equation}
where only $3$ independent variational parameters $\{\alpha,\xi,\zeta_{\alpha} \}$ are required in this case.

\begin{figure}[htbp!]
\centering
\includegraphics[width=0.8\linewidth]{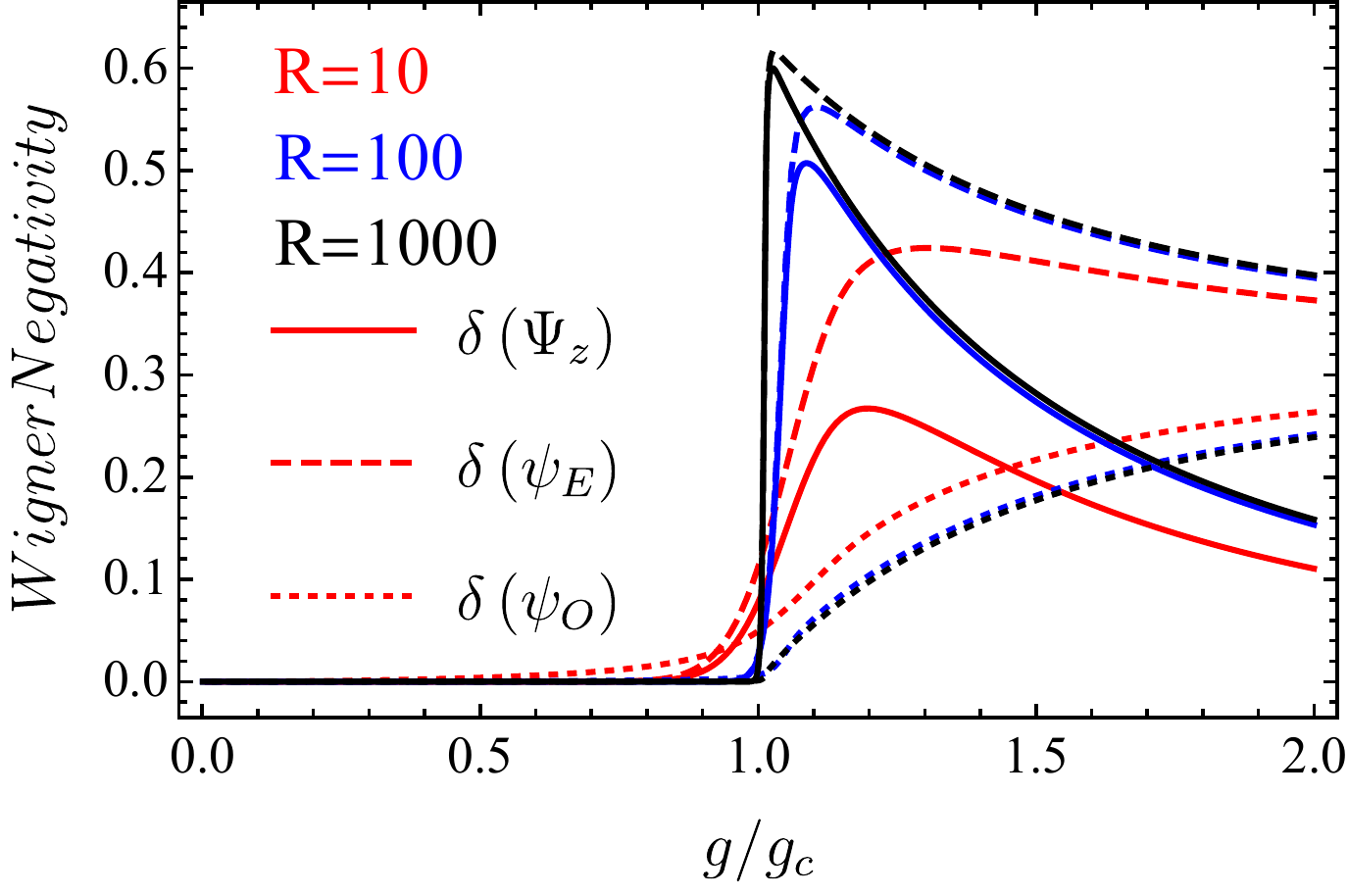}
\caption{The Wigner negativity by numerical integration of Eq. (\ref{WN}) as a function of the coupling strength $g/g_{c}$ for different ratio parameter $R$. The solid, dashed and dotted lines denote Wigner negativity for the Wigner functions of $\Psi_{z}$, $\psi_{E}$, and $\psi_{O}$, respectively, the red, blue and black colors correspond to  results of the ratio parameter $R=10$, $R=100$, and $R=1000$.}
\label{Wigner_Negativity}
\end{figure}

The Wigner negativity  is evaluated to measure nonclassicality  for the cavity states \cite{Kenfack2004JOB},  which is defined as
\begin{equation}\label{WN}
	\delta\left(\psi\right)=\iint \left[ \left|W_{\psi}\left(x,p\right) \right|-W_{\psi}\left(x,p\right) \right] \mathrm{d} x \mathrm{d} p.
\end{equation}
In Fig. \ref{Wigner_Negativity}, we calculate the Wigner negativity for  Wigner functions of the ground state  by numerical integration, the solid, dashed and dotted lines denote Wigner negativity for the Wigner functions $W_{T}$, $W_{z}^{E}$ and $W_{z}^{O}$ corresponding to the wave functions of the ground state $\Psi_{z}$, the even state $\psi_{E}$, and the odd state  $\psi_{O}$, respectively, the red, blue and black colors correspond to results of the ratio parameter $R=10$, $R=100$, and $R=1000$.
The quantities of  Wigner negativity are all equal to zero within vacuum state and squeezed vacuum state when the coupling strength $g/g_{c}\lesssim 1$.
For $g/g_{c}\gtrsim 1$ and larger ratio parameter $R$, $\delta\left(\Psi_{z}\right)$ and $\delta\left(\psi_{E}\right)$ increase more sharply around $g_{c}$, both reach their larger maximum when the coupling strength is slightly greater than the critical point $g_{c}$ and then start to decline,  while $\delta\left(\psi_{O}\right) $  keeps growing but is smaller as the coupling strength  increasing.
In addition, for the same $R$, $\delta\left(\psi_{E}\right)>\delta\left(\Psi_{z}\right)$ and $\delta\left(\psi_{E}\right)>\delta\left(\psi_{O}\right)$ for  $g/g_{c} > 1$, $\delta\left(\Psi_{z}\right)$  decreases faster than $\delta\left(\psi_{E}\right)$  due to $\delta\left(\psi_{O}\right)$  increasing with coupling strength. Further, $\delta\left(\psi_{O}\right)>\delta\left(\Psi_{z}\right)$ when the coupling strength reaches enough strong.

Combining with the above analysis, we can obtain that the Wigner negativity for ground state of the QRM is non-zero when the displacements $D_{\alpha,\beta}$ is non-zero, and when the ratio parameter $R$ is larger, the squeezing is larger, the values of $\delta\left(\Psi_{z}\right)$ and $\delta\left(\psi_{E}\right)$ are larger, while $\delta\left(\psi_{O}\right)$ is smaller.
To sum up, the presence or absence of the displacements is decisive for the occurrence of nonclassicality of the cavity states, while the squeezing only affects its extreme value.

The existence of generalized cat states in the QRM, including the odd cat states, even cat states and exotic semi-cat states with/without squeezing, will provide more possibilities for  quantum computing  \cite{Hacker2019NP,Bergmann2016PRA,Grimm2020Nature} and quantum information processing  \cite{Li2017PRL,Albert2016PRL,Sun2021PRL} based on the interaction of light and matter.

\subsection{\label{sec:level3B} The entangled state resulting from the emergence of spin-up state}
\begin{figure}[htp]
	\centering
	\includegraphics[width=0.8\linewidth]{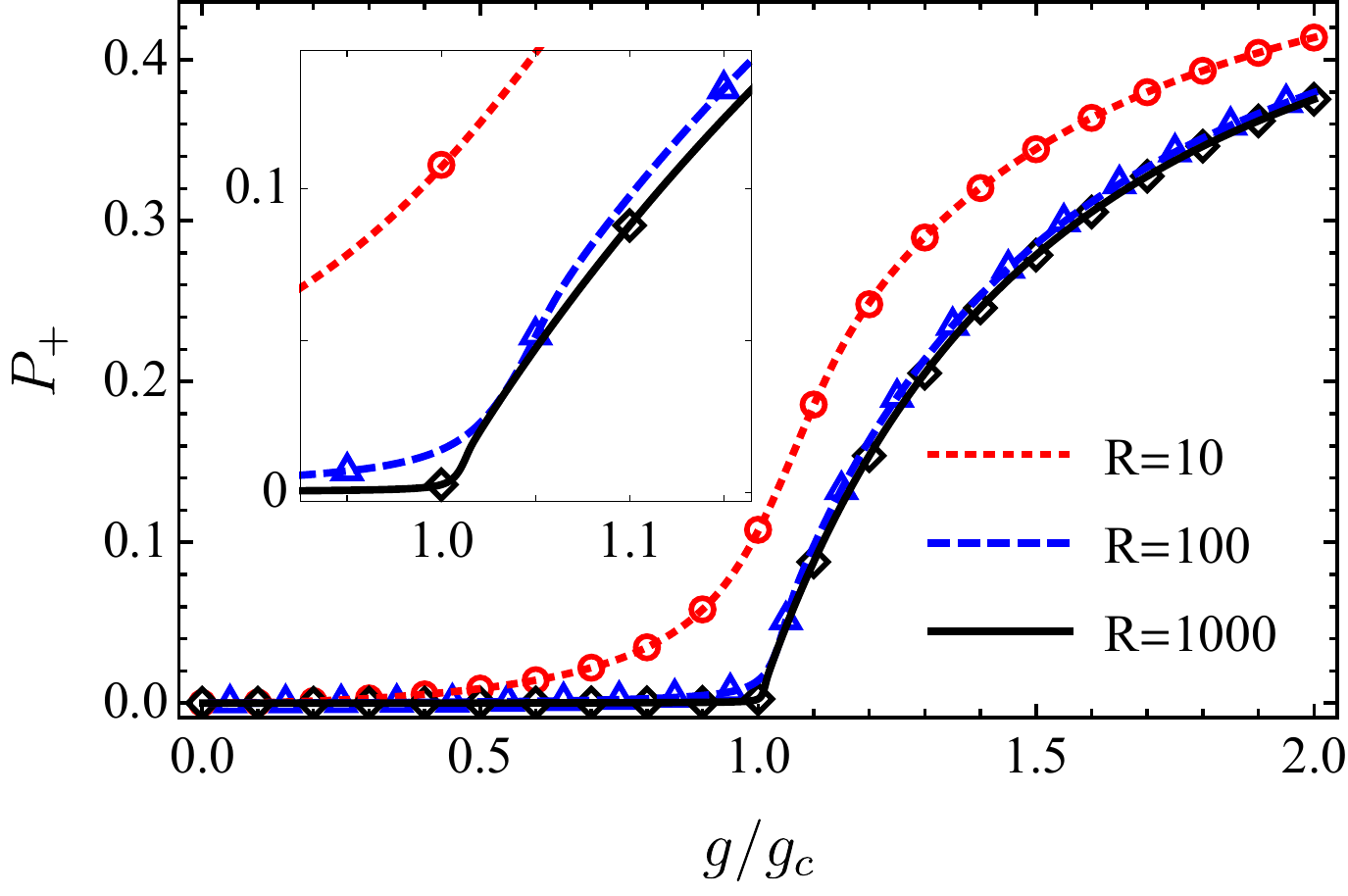}
	\caption{The probability  $P_{+}$  as function of the coupling strength $g/g_{c}$ for ground state of the QRM with different ratio parameter $R$.  The red dotted, blue dashed and black solid lines are obtained by using polaron picture for $R=10$, $100$ and $1000$, respectively.
		The corresponding symbols denote the numerically exact results.}
	\label{Ppmz}
\end{figure}

Note that the population of the spin-up state  $\left|\uparrow \right\rangle_{z}$  for the ground state is also used as the order parameters of the SPT in the QRM \cite{Hwang2015PRL,Cai2021NC}.
The probabilities of the states for the two-level system in $\sigma_{z}$ representation are given by
\begin{equation}\label{Ppm}
		P_{\pm} =\frac{1}{2}[\alpha^{2}(1\mp T^{\alpha}_{\alpha})+\beta^{2}(1 \mp T^{\beta}_{\beta})
\mp 2\alpha\beta T^{\alpha}_{\beta}(1/T^{\gamma}_{\gamma}\mp 1)],		
\end{equation}
where $\gamma=\sqrt{\alpha\beta}$,  $P_{+}= \left \langle \psi_{O}|\psi_{O}\right\rangle /4$ and $P_{-}=\left \langle \psi_{E}|\psi_{E}\right\rangle /4$ represent the probabilities of spin-up  state $\left|\uparrow\right\rangle_{z}$  and -down  state $\left|\downarrow\right\rangle_{z}$, respectively. The probability $P_{+}$ as function of the coupling strength for different $R$ is indicated in Fig. \ref{Ppmz}, and $P_{-}=1-P_{+}$.

\begin{figure}[htbp!]
	\centering
	\includegraphics[width=0.8\linewidth]{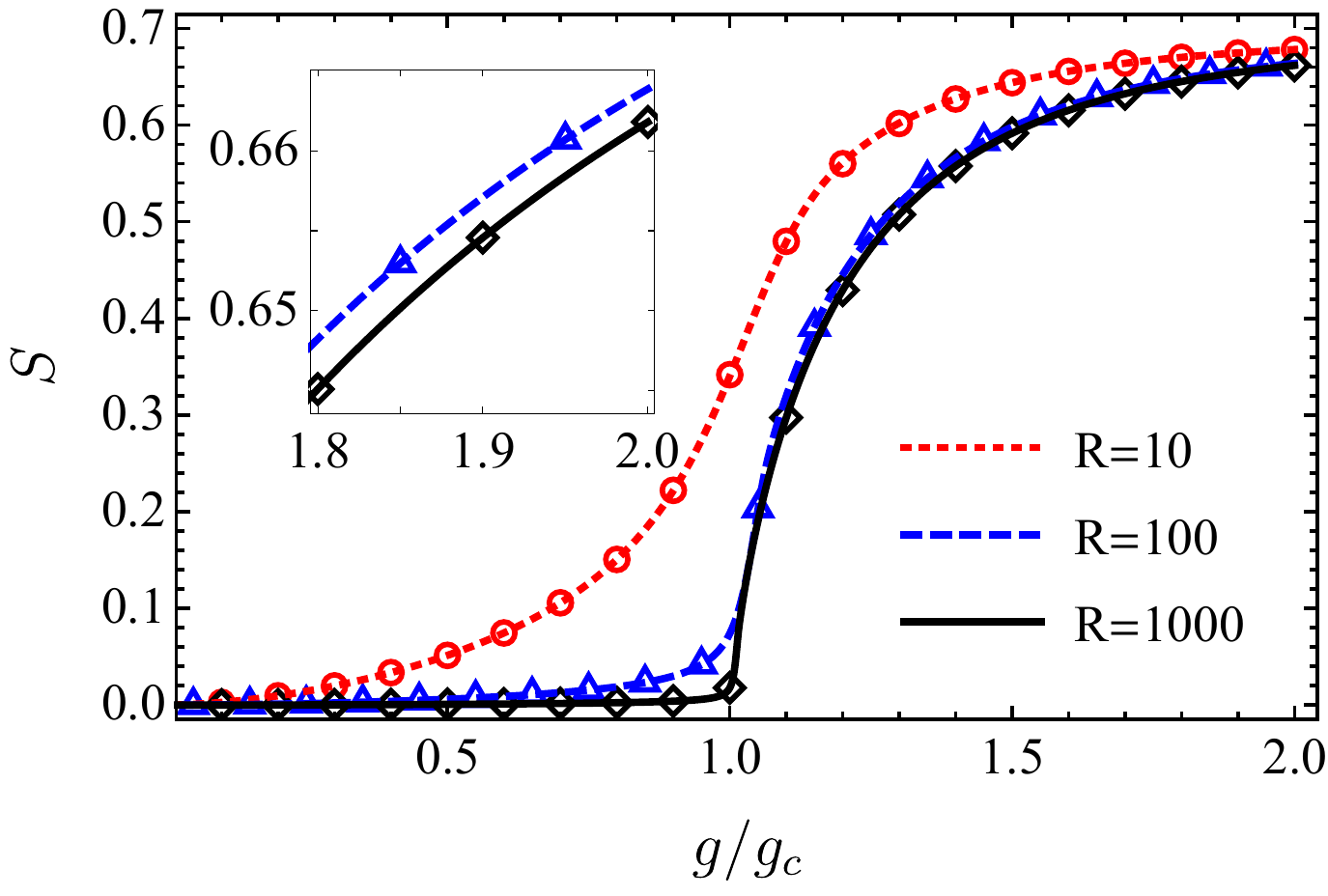}
	\caption{The Von Neumann entanglement entropy as function of the coupling strength $g/g_{c}$ for  the ground state of the QRM with different $R$. The red dotted, blue dashed and black solid lines denote the entanglement entropy by using the polaron picture for the ratio parameters $R=10$, $100$ and $1000$, respectively. The symbols are the numerically exact results as a benchmark.}
	\label{EE}
\end{figure}

To measure entanglement between the harmonic oscillator and the two-level system in the ground state of the QRM, we calculate  Von Neumann entanglement entropy of the spin state by tracing out the cavity degree of freedom
\begin{equation}
	S=-\left(P_{-}\log P_{-}+P_{+}\log P_{+}\right),
\end{equation}
which is good agreement with the numerical result as shown in Fig. \ref{EE}.
For the same coupling strength, the smaller ratio parameter $R$ is, the larger the entanglement entropy is.
When $R$ is large enough, the SPT occurs with increasing coupling strength, in the normal phase,  $P_{+} \rightarrow 0$, and the ground-state wave function $\Psi_{z}=\psi_{E}\left|\downarrow\right\rangle_{z}$, is a separable state, where the quantum oscillator is in the vacuum states and the two-level system is in the spin-down state $\left|\downarrow\right\rangle_{z}$; in the superradiant phase, the spin-up state $\left|\uparrow\right\rangle_{z}$ emerges (see Fig. \ref{Ppmz}), and the ground state becomes entangled state (\ref{WFz}), where the odd cat states $\psi_{O}$  and the even cat states $\psi_{E}$ entangled with the spin-up state $\left|\uparrow\right\rangle_{z}$ and -down state $\left|\downarrow\right\rangle_{z}$, respectively.

From Fig. \ref{EE}, it can be seen that the entanglement entropy increases more sharply from zero to a finite value  near the critical point for larger ratio parameter $R$, which can also be used as the order parameter of the superradiant phase transition \cite{Ashhab2013PRA,Liu2017PRL,Lin2017JPCS}.

\section{\label{sec:level4}Classification for the ground state and the processes of the SPT }
By analyzed the variational parameters and calculated the related physical quantities, it can be seen that the boundaries between different nonclassical states based on the squeezing and superradiance are obvious when the ratio parameter $R$ is relatively large, thus the different phase regions for the ground state of QRM within corresponding coupling strength intervals are distinguished as shown in Table. \ref{phase_regions}.
The squeezing region is in  the coupling strength interval of $0.5\lesssim g/g_{c}\lesssim1.5$, and the critical point $g_{c}$ divides the normal phase and superradiant phase. In the region of the normal phase, the ground state is the separable state, where the vacuum states of the quantum oscillator begin to be squeezed in the $p$ direction when $g/g_{c}\gtrsim0.5$ and becomes squeezed vacuum states while the two-level system is in the down state $\left|\downarrow\right\rangle_{z}$, the squeezing increases with the coupling strength and reaches a maximum near $g_{c}$. In the region of the superradiant phase,  the ground state becomes the entangled state, the squeezing starts to decrease and the displacements change from zero to a finite values in the superradiant phase region, correspondingly, the rapidly growing cat states in size with squeezing emerge, which consist of a large amount of the even cat states entangled with spin-down state and a small amount of the odd cat states entangled with the spin-up state in $\sigma_{z}$ representation. To further increase the coupling strength such that
$g/g_{c}\gtrsim1.5$,  the squeezing vanishes, and the novel semi-cat states are obtained in $\sigma_{x}$ representation due to the antipolaron with the small but nonzero weight. Hence, the ground state of the QRM with large ratio parameter $R$ is divided into four regions in the coupling strength based on the properties of nonclassical states:  (a) vacuum states (VS), (b) squeezed vacuum states (SVS), (c) squeezed cat states (SCS), and (d) cat states without squeezing (CSWS), the corresponding Wigner functions are shown in Fig. \ref{WFt}.

\begin{table}[htbp]
	\caption{The  nonclassical-state regions for ground state of the QRM within the SPT.}
	\label{phase_regions}
	\centering
	\begin{tabular}{@{}c|c|c|c}
		\hline\hline
		$g/g_{c}\lesssim0.5$        &       $0.5\lesssim g/g_{c}\lesssim1$             &  $1\lesssim g/g_{c}\lesssim 1.5$    &   $g/g_{c}\gtrsim 1.5$        \\ \hline
		\multicolumn{2}{c|}{\multirow{1}{*}{Normal phase}}       &  \multicolumn{2}{c}{\multirow{1}{*}{Superradiant phase}} \\ \hline
		\multicolumn{2}{c|}{\multirow{1}{*}{Separable state}}       &  \multicolumn{2}{c}{\multirow{1}{*}{Entangled state}} \\ \hline
		&  \multicolumn{2}{c|}{\multirow{1}{*}{Squeezed region}}  &        \\ \hline
		VS &    SVS          &  SCS   &  CSWS       \\ \hline
		\hline
	\end{tabular}%
\end{table}

\begin{figure}[htbp]
	\centering
	\includegraphics[width=\linewidth]{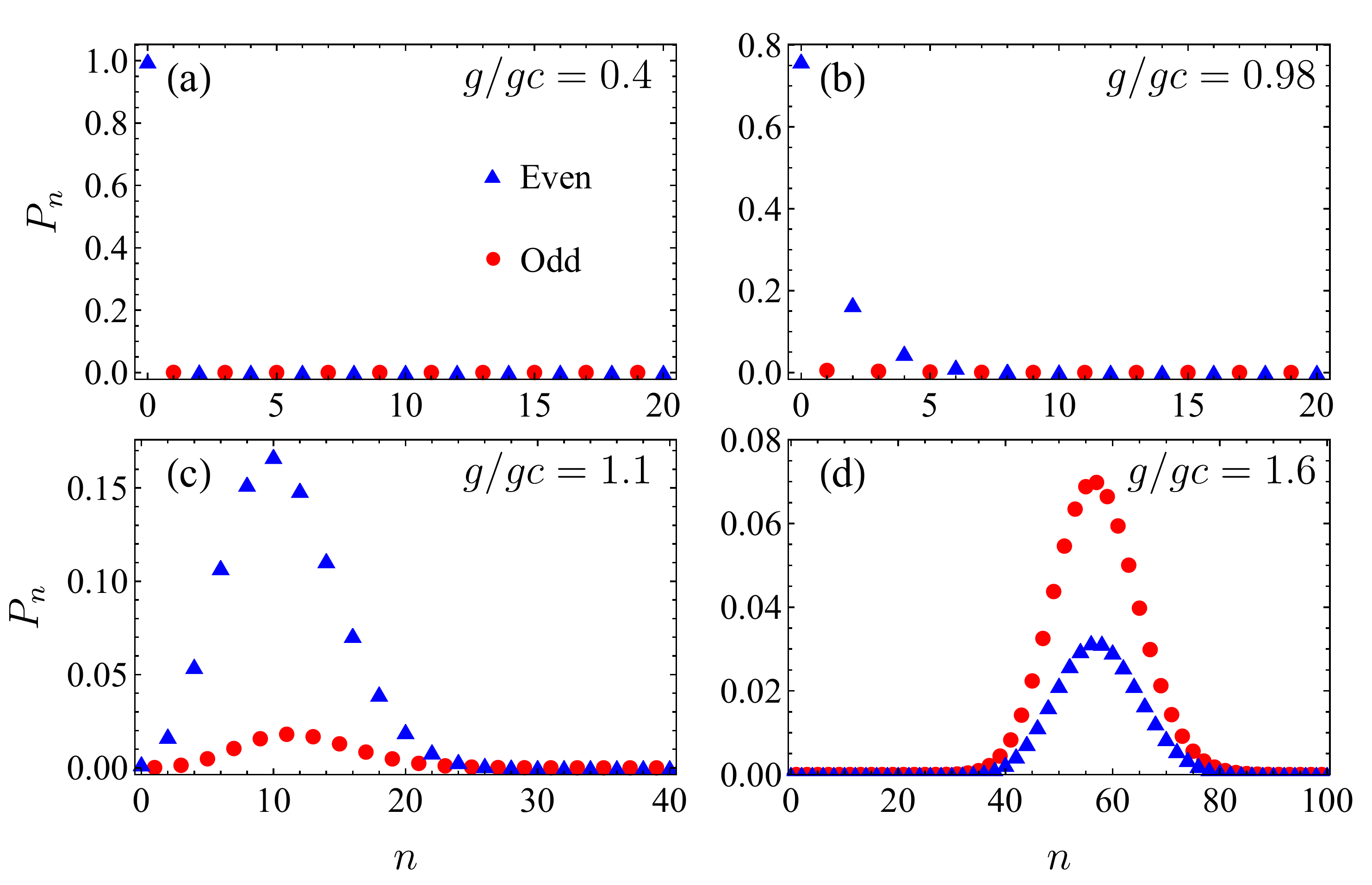}
	\caption{The photon number distribution in Fock space for (a) the vacuum states, (b) the squeezed vacuum states,  (c) the squeezed cat states, and (d) the cat states without squeezing contained in the ground state.  The red points and blue triangles denote the photon distribution in Fock basis with odd and even photon numbers, respectively. The parameters are the same as in Fig. \ref{WFt}.}
	\label{PND}
\end{figure}

Furthermore, we calculate the photon number distribution in Fock space for the ground state to figure out the processes of the SPT as shown in Fig. \ref{PND}. It can be seen that: the number of photons is all in $\left|0\right\rangle$ for the vacuum states;
the photon probability distribution for the squeezed vacuum states is only existing for the even photon numbers and vanishing for all odd photon numbers \cite{Gerry2005IntroductoryQO}, which can be fitted by the Poissonian-like statistics; and the photon population in odd photon numbers arises for the squeezed odd cat states and odd cat states without squeezing, together with the photon population in even photon numbers for the even counterparts, corresponding to the statistics of Gaussian unitary ensemble (GUE)\cite{Yang2022CharacterizingSP}.
Notably, the mean number of photons of the squeezed vacuum states is zero, and when the squeeze decreases after the critical point and the displacements of the two polarons occur, the mean number of photons increases dramatically.
In contrast to the case of the smaller  $R$, there is no squeezing process with the coupling strength increasing, so the mean number of photons in the ground state increases continuously.

\begin{figure}[htbp!]
	\centering
	\includegraphics[width=0.68\linewidth]{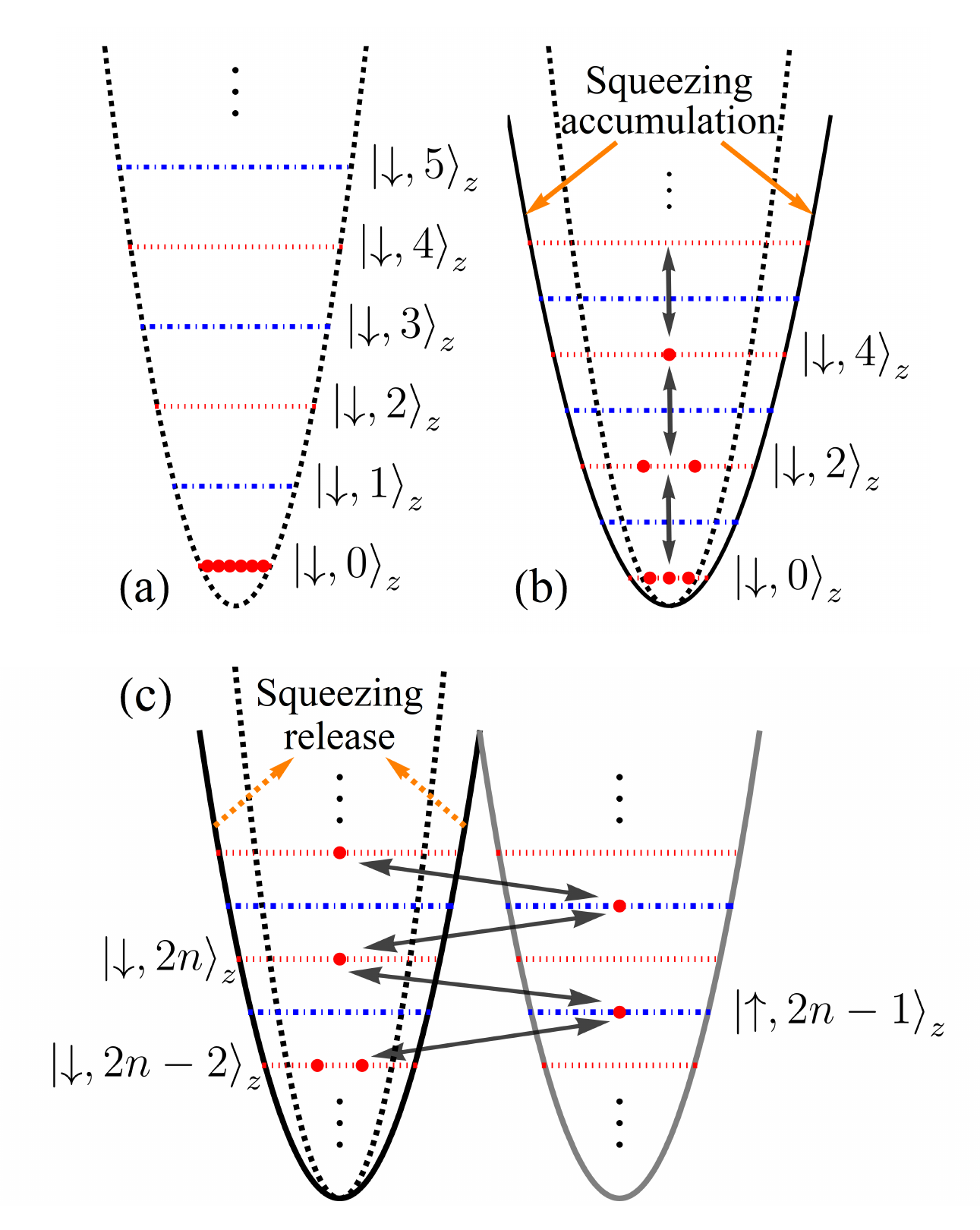}
	\caption{Schematic diagram for the ground-state photon number distribution in Fock space for (a) the vacuum states with only photon population for $\left|0\right\rangle$ in branch of the spin-down state $\left|\downarrow \right \rangle_{z}$, (b) a squeezing accumulation leading to the squeezed vacuum states with Poissonian-like photon distribution for the even photon numbers in the branch of  $\left|\downarrow \right \rangle_{z}$,  (c)  a squeezing release leading to the emergence of the squeezed odd cat states with GUE-like photon distribution for the odd photon numbers in branch the spin-up state of  $\left|\uparrow \right \rangle_{z}$, while the GUE-like photon distribution for even photon numbers forms the squeezed even cat states in the branch of  $\left|\downarrow \right \rangle_{z}$. The dashed lines denote the potential of the harmonic oscillator without squeezing, the  black and gray solid lines denote the potentials of the harmonic oscillator with squeezing corresponding to the branches of the spin-down  and spin-up states, respectively.}
	\label{SD}
\end{figure}

In this point of view, combined with the state of the two-level system, we can understand processes of the SPT with increasing coupling strength as transfer of the photon number population from  $\left|\downarrow,0 \right\rangle_{z}$ to the distribution of only the even photon numbers with Poissonian-like statistics leading to the accumulation of squeezing before the critical point (see Fig. \ref{SD}(a,b)), corresponding to $\left|\downarrow,0 \right\rangle_{z}\leftrightarrow\left|\downarrow,2 \right\rangle_{z}\leftrightarrow \cdots$, and the squeezing reaches the maximum near the critical point; then emergence of the spin-up state and GUE-like photon population within the odd photon numbers leading to  the release of the squeezing after the critical point is accompanied by a sharp increase in the displacements after the critical point (see Fig. \ref{SD}(c)), i.e. $\cdots\leftrightarrow\left|\uparrow,2n-2\right\rangle_{z}\leftrightarrow\left|\downarrow,2n-1 \right\rangle_{z}\leftrightarrow\left|\uparrow,2n \right\rangle_{z}\leftrightarrow \cdots$, and forms the ground-state wave function in Eq. (\ref{WFz}) which consists of the cat states entangled with the spin states.

\section{\label{Conclusion}Conclusion}
We study the nonclassical states contained in ground state of the QRM using polaron picture, where the variational wave function consists of a polaron and an antipolaron, which have the same squeezing parameter and are distinguished by different displacements and weights.
Further, we derive analytical expressions of  Wigner functions for the ground states to characterize the nonclassical states, and calculate Wigner negativity  and  Von Neumann entanglement entropy to measure  nonclassicality and entanglement, respectively, which are important for the modern quantum technologies.

A squeezed region within all parameter ranges is  distinguished, which corresponds to the region around the critical coupling strength of the SPT in the model with significant ratio parameters, before and after the critical point are the squeezed vacuum states and the squeezed cat states, respectively.
And a pair of novel semi-cat states caused by the necessary antipolaron with small but nonzero weight when the coupling strength is strong enough is revealed.
We find that the $x$-type SPT is dominated by the cat states with displacements of the same size increasing from zero to a finite value when the coupling strength goes through the critical point, while the ground state changes from a separated state to an entangled state.
And we give the interval of coupling strengths  corresponding to the nonclassical states based on the squeezing and the superradiance.
Moreover, by combining  the photon number distribution in Fock space for the ground state of the QRM, we clearly indicate the process of SPT, that is, the system sequentially goes through the vacuum states, the squeezed vacuum states, the squeezed cat states, and the cat states without squeezing as the coupling strength increases.

\begin{acknowledgments}
We thank G. Liu and F.-Z. Chen for helpful discussions.
We acknowledge funding from the National Key Research and Development Program of China (Grant No. 2022YFA1402704) and the National Natural Science Foundation of China (Grant No. 12047501 and No. 11834005).
\end{acknowledgments}

\bibliographystyle{apsrev4-1}
\bibliography{ref}

\begin{thebibliography}{45}%
\makeatletter
\providecommand \@ifxundefined [1]{%
 \@ifx{#1\undefined}
}%
\providecommand \@ifnum [1]{%
 \ifnum #1\expandafter \@firstoftwo
 \else \expandafter \@secondoftwo
 \fi
}%
\providecommand \@ifx [1]{%
 \ifx #1\expandafter \@firstoftwo
 \else \expandafter \@secondoftwo
 \fi
}%
\providecommand \natexlab [1]{#1}%
\providecommand \enquote  [1]{``#1''}%
\providecommand \bibnamefont  [1]{#1}%
\providecommand \bibfnamefont [1]{#1}%
\providecommand \citenamefont [1]{#1}%
\providecommand \href@noop [0]{\@secondoftwo}%
\providecommand \href [0]{\begingroup \@sanitize@url \@href}%
\providecommand \@href[1]{\@@startlink{#1}\@@href}%
\providecommand \@@href[1]{\endgroup#1\@@endlink}%
\providecommand \@sanitize@url [0]{\catcode `\\12\catcode `\$12\catcode
  `\&12\catcode `\#12\catcode `\^12\catcode `\_12\catcode `\%12\relax}%
\providecommand \@@startlink[1]{}%
\providecommand \@@endlink[0]{}%
\providecommand \url  [0]{\begingroup\@sanitize@url \@url }%
\providecommand \@url [1]{\endgroup\@href {#1}{\urlprefix }}%
\providecommand \urlprefix  [0]{URL }%
\providecommand \Eprint [0]{\href }%
\providecommand \doibase [0]{http://dx.doi.org/}%
\providecommand \selectlanguage [0]{\@gobble}%
\providecommand \bibinfo  [0]{\@secondoftwo}%
\providecommand \bibfield  [0]{\@secondoftwo}%
\providecommand \translation [1]{[#1]}%
\providecommand \BibitemOpen [0]{}%
\providecommand \bibitemStop [0]{}%
\providecommand \bibitemNoStop [0]{.\EOS\space}%
\providecommand \EOS [0]{\spacefactor3000\relax}%
\providecommand \BibitemShut  [1]{\csname bibitem#1\endcsname}%
\let\auto@bib@innerbib\@empty
\bibitem [{\citenamefont {Rabi}(1937)}]{Rabi1937PR}%
  \BibitemOpen
  \bibfield  {author} {\bibinfo {author} {\bibfnamefont {I.~I.}\ \bibnamefont
  {Rabi}},\ }\href {\doibase 10.1103/PhysRev.51.652} {\bibfield  {journal}
  {\bibinfo  {journal} {Phys. Rev.}\ }\textbf {\bibinfo {volume} {51}},\
  \bibinfo {pages} {652} (\bibinfo {year} {1937})}\BibitemShut {NoStop}%
\bibitem [{\citenamefont {Forn-D\'{\i}az}\ \emph {et~al.}(2019)\citenamefont
  {Forn-D\'{\i}az}, \citenamefont {Lamata}, \citenamefont {Rico}, \citenamefont
  {Kono},\ and\ \citenamefont {Solano}}]{Forn2019RMP}%
  \BibitemOpen
  \bibfield  {author} {\bibinfo {author} {\bibfnamefont {P.}~\bibnamefont
  {Forn-D\'{\i}az}}, \bibinfo {author} {\bibfnamefont {L.}~\bibnamefont
  {Lamata}}, \bibinfo {author} {\bibfnamefont {E.}~\bibnamefont {Rico}},
  \bibinfo {author} {\bibfnamefont {J.}~\bibnamefont {Kono}}, \ and\ \bibinfo
  {author} {\bibfnamefont {E.}~\bibnamefont {Solano}},\ }\href {\doibase
  10.1103/RevModPhys.91.025005} {\bibfield  {journal} {\bibinfo  {journal}
  {Rev. Mod. Phys.}\ }\textbf {\bibinfo {volume} {91}},\ \bibinfo {pages}
  {025005} (\bibinfo {year} {2019})}\BibitemShut {NoStop}%
\bibitem [{\citenamefont {Li}\ and\ \citenamefont
  {Batchelor}(2021)}]{Li2021PRA}%
  \BibitemOpen
  \bibfield  {author} {\bibinfo {author} {\bibfnamefont {Z.-M.}\ \bibnamefont
  {Li}}\ and\ \bibinfo {author} {\bibfnamefont {M.~T.}\ \bibnamefont
  {Batchelor}},\ }\href {\doibase 10.1103/PhysRevA.104.033712} {\bibfield
  {journal} {\bibinfo  {journal} {Phys. Rev. A}\ }\textbf {\bibinfo {volume}
  {104}},\ \bibinfo {pages} {033712} (\bibinfo {year} {2021})}\BibitemShut
  {NoStop}%
\bibitem [{\citenamefont {{Jaynes}}\ and\ \citenamefont
  {{Cummings}}(1963)}]{Jaynes1963PIEEE}%
  \BibitemOpen
  \bibfield  {author} {\bibinfo {author} {\bibfnamefont {E.~T.}\ \bibnamefont
  {{Jaynes}}}\ and\ \bibinfo {author} {\bibfnamefont {F.~W.}\ \bibnamefont
  {{Cummings}}},\ }\href {\doibase 10.1109/PROC.1963.1664} {\bibfield
  {journal} {\bibinfo  {journal} {Proc. IEEE}\ }\textbf {\bibinfo {volume}
  {51}},\ \bibinfo {pages} {89} (\bibinfo {year} {1963})}\BibitemShut {NoStop}%
\bibitem [{\citenamefont {Brune}\ \emph {et~al.}(1996)\citenamefont {Brune},
  \citenamefont {Schmidt-Kaler}, \citenamefont {Maali}, \citenamefont {Dreyer},
  \citenamefont {Hagley}, \citenamefont {Raimond},\ and\ \citenamefont
  {Haroche}}]{Brune1996PRL}%
  \BibitemOpen
  \bibfield  {author} {\bibinfo {author} {\bibfnamefont {M.}~\bibnamefont
  {Brune}}, \bibinfo {author} {\bibfnamefont {F.}~\bibnamefont
  {Schmidt-Kaler}}, \bibinfo {author} {\bibfnamefont {A.}~\bibnamefont
  {Maali}}, \bibinfo {author} {\bibfnamefont {J.}~\bibnamefont {Dreyer}},
  \bibinfo {author} {\bibfnamefont {E.}~\bibnamefont {Hagley}}, \bibinfo
  {author} {\bibfnamefont {J.~M.}\ \bibnamefont {Raimond}}, \ and\ \bibinfo
  {author} {\bibfnamefont {S.}~\bibnamefont {Haroche}},\ }\href {\doibase
  10.1103/PhysRevLett.76.1800} {\bibfield  {journal} {\bibinfo  {journal}
  {Phys. Rev. Lett.}\ }\textbf {\bibinfo {volume} {76}},\ \bibinfo {pages}
  {1800} (\bibinfo {year} {1996})}\BibitemShut {NoStop}%
\bibitem [{\citenamefont {Thompson}\ \emph {et~al.}(1992)\citenamefont
  {Thompson}, \citenamefont {Rempe},\ and\ \citenamefont
  {Kimble}}]{Thompson1992PRL}%
  \BibitemOpen
  \bibfield  {author} {\bibinfo {author} {\bibfnamefont {R.~J.}\ \bibnamefont
  {Thompson}}, \bibinfo {author} {\bibfnamefont {G.}~\bibnamefont {Rempe}}, \
  and\ \bibinfo {author} {\bibfnamefont {H.~J.}\ \bibnamefont {Kimble}},\
  }\href {\doibase 10.1103/PhysRevLett.68.1132} {\bibfield  {journal} {\bibinfo
   {journal} {Phys. Rev. Lett.}\ }\textbf {\bibinfo {volume} {68}},\ \bibinfo
  {pages} {1132} (\bibinfo {year} {1992})}\BibitemShut {NoStop}%
\bibitem [{\citenamefont {Anappara}\ \emph {et~al.}(2009)\citenamefont
  {Anappara}, \citenamefont {De~Liberato}, \citenamefont {Tredicucci},
  \citenamefont {Ciuti}, \citenamefont {Biasiol}, \citenamefont {Sorba},\ and\
  \citenamefont {Beltram}}]{Anappara2009PRB}%
  \BibitemOpen
  \bibfield  {author} {\bibinfo {author} {\bibfnamefont {A.~A.}\ \bibnamefont
  {Anappara}}, \bibinfo {author} {\bibfnamefont {S.}~\bibnamefont
  {De~Liberato}}, \bibinfo {author} {\bibfnamefont {A.}~\bibnamefont
  {Tredicucci}}, \bibinfo {author} {\bibfnamefont {C.}~\bibnamefont {Ciuti}},
  \bibinfo {author} {\bibfnamefont {G.}~\bibnamefont {Biasiol}}, \bibinfo
  {author} {\bibfnamefont {L.}~\bibnamefont {Sorba}}, \ and\ \bibinfo {author}
  {\bibfnamefont {F.}~\bibnamefont {Beltram}},\ }\href {\doibase
  10.1103/PhysRevB.79.201303} {\bibfield  {journal} {\bibinfo  {journal} {Phys.
  Rev. B}\ }\textbf {\bibinfo {volume} {79}},\ \bibinfo {pages} {201303}
  (\bibinfo {year} {2009})}\BibitemShut {NoStop}%
\bibitem [{\citenamefont {Niemczyk}\ \emph {et~al.}(2010)\citenamefont
  {Niemczyk}, \citenamefont {Deppe}, \citenamefont {Huebl}, \citenamefont
  {Menzel}, \citenamefont {Hocke}, \citenamefont {Schwarz}, \citenamefont
  {Garcia-Ripoll}, \citenamefont {Zueco}, \citenamefont {Hümmer},
  \citenamefont {Solano}, \citenamefont {Marx},\ and\ \citenamefont
  {Gross}}]{Niemczyk2010NP}%
  \BibitemOpen
  \bibfield  {author} {\bibinfo {author} {\bibfnamefont {T.}~\bibnamefont
  {Niemczyk}}, \bibinfo {author} {\bibfnamefont {F.}~\bibnamefont {Deppe}},
  \bibinfo {author} {\bibfnamefont {H.}~\bibnamefont {Huebl}}, \bibinfo
  {author} {\bibfnamefont {E.~P.}\ \bibnamefont {Menzel}}, \bibinfo {author}
  {\bibfnamefont {F.}~\bibnamefont {Hocke}}, \bibinfo {author} {\bibfnamefont
  {M.~J.}\ \bibnamefont {Schwarz}}, \bibinfo {author} {\bibfnamefont {J.~J.}\
  \bibnamefont {Garcia-Ripoll}}, \bibinfo {author} {\bibfnamefont
  {D.}~\bibnamefont {Zueco}}, \bibinfo {author} {\bibfnamefont
  {T.}~\bibnamefont {Hümmer}}, \bibinfo {author} {\bibfnamefont
  {E.}~\bibnamefont {Solano}}, \bibinfo {author} {\bibfnamefont
  {A.}~\bibnamefont {Marx}}, \ and\ \bibinfo {author} {\bibfnamefont
  {R.}~\bibnamefont {Gross}},\ }\href {\doibase 10.1038/nphys1730} {\bibfield
  {journal} {\bibinfo  {journal} {Nat. Phys.}\ }\textbf {\bibinfo {volume}
  {6}},\ \bibinfo {pages} {772} (\bibinfo {year} {2010})}\BibitemShut {NoStop}%
\bibitem [{\citenamefont {Forn-D\'{\i}az}\ \emph {et~al.}(2017)\citenamefont
  {Forn-D\'{\i}az}, \citenamefont {Garc\'{\i}a-Ripoll}, \citenamefont
  {Peropadre}, \citenamefont {Orgiazzi}, \citenamefont {Yurtalan},
  \citenamefont {Belyansky}, \citenamefont {Wilson},\ and\ \citenamefont
  {Lupascu}}]{FornDiaz2017NP}%
  \BibitemOpen
  \bibfield  {author} {\bibinfo {author} {\bibfnamefont {P.}~\bibnamefont
  {Forn-D\'{\i}az}}, \bibinfo {author} {\bibfnamefont {J.~Â.}\ \bibnamefont
  {Garc\'{\i}a-Ripoll}}, \bibinfo {author} {\bibfnamefont {B.}~\bibnamefont
  {Peropadre}}, \bibinfo {author} {\bibfnamefont {J.~L.}\ \bibnamefont
  {Orgiazzi}}, \bibinfo {author} {\bibfnamefont {M.~Â.}\ \bibnamefont
  {Yurtalan}}, \bibinfo {author} {\bibfnamefont {R.}~\bibnamefont {Belyansky}},
  \bibinfo {author} {\bibfnamefont {C.~Â.}\ \bibnamefont {Wilson}}, \ and\
  \bibinfo {author} {\bibfnamefont {A.}~\bibnamefont {Lupascu}},\ }\href
  {\doibase 10.1038/nphys3905} {\bibfield  {journal} {\bibinfo  {journal} {Nat.
  Phys.}\ }\textbf {\bibinfo {volume} {13}},\ \bibinfo {pages} {39} (\bibinfo
  {year} {2017})}\BibitemShut {NoStop}%
\bibitem [{\citenamefont {De~Liberato}(2014)}]{Liberato2014PRL}%
  \BibitemOpen
  \bibfield  {author} {\bibinfo {author} {\bibfnamefont {S.}~\bibnamefont
  {De~Liberato}},\ }\href {\doibase 10.1103/PhysRevLett.112.016401} {\bibfield
  {journal} {\bibinfo  {journal} {Phys. Rev. Lett.}\ }\textbf {\bibinfo
  {volume} {112}},\ \bibinfo {pages} {016401} (\bibinfo {year}
  {2014})}\BibitemShut {NoStop}%
\bibitem [{\citenamefont {Langford}\ \emph {et~al.}(2017)\citenamefont
  {Langford}, \citenamefont {Sagastizabal}, \citenamefont {Kounalakis},
  \citenamefont {Dickel}, \citenamefont {Bruno}, \citenamefont {Luthi},
  \citenamefont {Thoen}, \citenamefont {Endo},\ and\ \citenamefont
  {DiCarlo}}]{Langford2017NC}%
  \BibitemOpen
  \bibfield  {author} {\bibinfo {author} {\bibfnamefont {N.~K.}\ \bibnamefont
  {Langford}}, \bibinfo {author} {\bibfnamefont {R.}~\bibnamefont
  {Sagastizabal}}, \bibinfo {author} {\bibfnamefont {M.}~\bibnamefont
  {Kounalakis}}, \bibinfo {author} {\bibfnamefont {C.}~\bibnamefont {Dickel}},
  \bibinfo {author} {\bibfnamefont {A.}~\bibnamefont {Bruno}}, \bibinfo
  {author} {\bibfnamefont {F.}~\bibnamefont {Luthi}}, \bibinfo {author}
  {\bibfnamefont {D.~J.}\ \bibnamefont {Thoen}}, \bibinfo {author}
  {\bibfnamefont {A.}~\bibnamefont {Endo}}, \ and\ \bibinfo {author}
  {\bibfnamefont {L.}~\bibnamefont {DiCarlo}},\ }\href {\doibase
  10.1038/s41467-017-01061-x} {\bibfield  {journal} {\bibinfo  {journal} {Nat.
  Commun.}\ }\textbf {\bibinfo {volume} {8}},\ \bibinfo {pages} {1715}
  (\bibinfo {year} {2017})}\BibitemShut {NoStop}%
\bibitem [{\citenamefont {Yoshihara}\ \emph {et~al.}(2017)\citenamefont
  {Yoshihara}, \citenamefont {Fuse}, \citenamefont {Ashhab}, \citenamefont
  {Kakuyanagi}, \citenamefont {Saito},\ and\ \citenamefont
  {Semba}}]{Yoshihara2017NP}%
  \BibitemOpen
  \bibfield  {author} {\bibinfo {author} {\bibfnamefont {F.}~\bibnamefont
  {Yoshihara}}, \bibinfo {author} {\bibfnamefont {T.}~\bibnamefont {Fuse}},
  \bibinfo {author} {\bibfnamefont {S.}~\bibnamefont {Ashhab}}, \bibinfo
  {author} {\bibfnamefont {K.}~\bibnamefont {Kakuyanagi}}, \bibinfo {author}
  {\bibfnamefont {S.}~\bibnamefont {Saito}}, \ and\ \bibinfo {author}
  {\bibfnamefont {K.}~\bibnamefont {Semba}},\ }\href {\doibase
  10.1038/nphys3906} {\bibfield  {journal} {\bibinfo  {journal} {Nat. Phys.}\
  }\textbf {\bibinfo {volume} {13}},\ \bibinfo {pages} {44} (\bibinfo {year}
  {2017})}\BibitemShut {NoStop}%
\bibitem [{\citenamefont {Frisk~Kockum}\ \emph {et~al.}(2019)\citenamefont
  {Frisk~Kockum}, \citenamefont {Miranowicz}, \citenamefont {De~Liberato},
  \citenamefont {Savasta},\ and\ \citenamefont {Nori}}]{Frisk2019NRP}%
  \BibitemOpen
  \bibfield  {author} {\bibinfo {author} {\bibfnamefont {A.}~\bibnamefont
  {Frisk~Kockum}}, \bibinfo {author} {\bibfnamefont {A.}~\bibnamefont
  {Miranowicz}}, \bibinfo {author} {\bibfnamefont {S.}~\bibnamefont
  {De~Liberato}}, \bibinfo {author} {\bibfnamefont {S.}~\bibnamefont
  {Savasta}}, \ and\ \bibinfo {author} {\bibfnamefont {F.}~\bibnamefont
  {Nori}},\ }\href {\doibase 10.1038/s42254-018-0006-2} {\bibfield  {journal}
  {\bibinfo  {journal} {Nat. Rev. Phys.}\ }\textbf {\bibinfo {volume} {1}},\
  \bibinfo {pages} {19} (\bibinfo {year} {2019})}\BibitemShut {NoStop}%
\bibitem [{\citenamefont {Rossatto}\ \emph {et~al.}(2017)\citenamefont
  {Rossatto}, \citenamefont {Villas-Bôas}, \citenamefont {Sanz},\ and\
  \citenamefont {Solano}}]{Rossatto2017PRA}%
  \BibitemOpen
  \bibfield  {author} {\bibinfo {author} {\bibfnamefont {D.~Z.}\ \bibnamefont
  {Rossatto}}, \bibinfo {author} {\bibfnamefont {C.~J.}\ \bibnamefont
  {Villas-Bôas}}, \bibinfo {author} {\bibfnamefont {M.}~\bibnamefont {Sanz}},
  \ and\ \bibinfo {author} {\bibfnamefont {E.}~\bibnamefont {Solano}},\ }\href
  {\doibase 10.1103/PhysRevA.96.013849} {\bibfield  {journal} {\bibinfo
  {journal} {Physical Review A}\ }\textbf {\bibinfo {volume} {96}},\ \bibinfo
  {pages} {013849} (\bibinfo {year} {2017})}\BibitemShut {NoStop}%
\bibitem [{\citenamefont {Hwang}\ \emph {et~al.}(2015)\citenamefont {Hwang},
  \citenamefont {Puebla},\ and\ \citenamefont {Plenio}}]{Hwang2015PRL}%
  \BibitemOpen
  \bibfield  {author} {\bibinfo {author} {\bibfnamefont {M.-J.}\ \bibnamefont
  {Hwang}}, \bibinfo {author} {\bibfnamefont {R.}~\bibnamefont {Puebla}}, \
  and\ \bibinfo {author} {\bibfnamefont {M.~B.}\ \bibnamefont {Plenio}},\
  }\href {\doibase 10.1103/PhysRevLett.115.180404} {\bibfield  {journal}
  {\bibinfo  {journal} {Phys. Rev. Lett.}\ }\textbf {\bibinfo {volume} {115}},\
  \bibinfo {pages} {180404} (\bibinfo {year} {2015})}\BibitemShut {NoStop}%
\bibitem [{\citenamefont {Liu}\ \emph {et~al.}(2017)\citenamefont {Liu},
  \citenamefont {Chesi}, \citenamefont {Ying}, \citenamefont {Chen},
  \citenamefont {Luo},\ and\ \citenamefont {Lin}}]{Liu2017PRL}%
  \BibitemOpen
  \bibfield  {author} {\bibinfo {author} {\bibfnamefont {M.}~\bibnamefont
  {Liu}}, \bibinfo {author} {\bibfnamefont {S.}~\bibnamefont {Chesi}}, \bibinfo
  {author} {\bibfnamefont {Z.-J.}\ \bibnamefont {Ying}}, \bibinfo {author}
  {\bibfnamefont {X.}~\bibnamefont {Chen}}, \bibinfo {author} {\bibfnamefont
  {H.-G.}\ \bibnamefont {Luo}}, \ and\ \bibinfo {author} {\bibfnamefont
  {H.-Q.}\ \bibnamefont {Lin}},\ }\href {\doibase
  10.1103/PhysRevLett.119.220601} {\bibfield  {journal} {\bibinfo  {journal}
  {Phys. Rev. Lett.}\ }\textbf {\bibinfo {volume} {119}},\ \bibinfo {pages}
  {220601} (\bibinfo {year} {2017})}\BibitemShut {NoStop}%
\bibitem [{\citenamefont {Cai}\ \emph {et~al.}(2021)\citenamefont {Cai},
  \citenamefont {Liu}, \citenamefont {Zhao}, \citenamefont {Wu}, \citenamefont
  {Mei}, \citenamefont {Jiang}, \citenamefont {He}, \citenamefont {Zhang},
  \citenamefont {Zhou},\ and\ \citenamefont {Duan}}]{Cai2021NC}%
  \BibitemOpen
  \bibfield  {author} {\bibinfo {author} {\bibfnamefont {M.~L.}\ \bibnamefont
  {Cai}}, \bibinfo {author} {\bibfnamefont {Z.~D.}\ \bibnamefont {Liu}},
  \bibinfo {author} {\bibfnamefont {W.~D.}\ \bibnamefont {Zhao}}, \bibinfo
  {author} {\bibfnamefont {Y.~K.}\ \bibnamefont {Wu}}, \bibinfo {author}
  {\bibfnamefont {Q.~X.}\ \bibnamefont {Mei}}, \bibinfo {author} {\bibfnamefont
  {Y.}~\bibnamefont {Jiang}}, \bibinfo {author} {\bibfnamefont
  {L.}~\bibnamefont {He}}, \bibinfo {author} {\bibfnamefont {X.}~\bibnamefont
  {Zhang}}, \bibinfo {author} {\bibfnamefont {Z.~C.}\ \bibnamefont {Zhou}}, \
  and\ \bibinfo {author} {\bibfnamefont {L.~M.}\ \bibnamefont {Duan}},\ }\href
  {\doibase 10.1038/s41467-021-21425-8} {\bibfield  {journal} {\bibinfo
  {journal} {Nat. Commun.}\ }\textbf {\bibinfo {volume} {12}},\ \bibinfo
  {pages} {1126} (\bibinfo {year} {2021})}\BibitemShut {NoStop}%
\bibitem [{\citenamefont {Cai}\ \emph {et~al.}(2022)\citenamefont {Cai},
  \citenamefont {Liu}, \citenamefont {Jiang}, \citenamefont {Wu}, \citenamefont
  {Mei}, \citenamefont {Zhao}, \citenamefont {He}, \citenamefont {Zhang},
  \citenamefont {Zhou},\ and\ \citenamefont {Duan}}]{Cai2022CPL}%
  \BibitemOpen
  \bibfield  {author} {\bibinfo {author} {\bibfnamefont {M.~L.}\ \bibnamefont
  {Cai}}, \bibinfo {author} {\bibfnamefont {Z.~D.}\ \bibnamefont {Liu}},
  \bibinfo {author} {\bibfnamefont {Y.}~\bibnamefont {Jiang}}, \bibinfo
  {author} {\bibfnamefont {Y.~K.}\ \bibnamefont {Wu}}, \bibinfo {author}
  {\bibfnamefont {Q.~X.}\ \bibnamefont {Mei}}, \bibinfo {author} {\bibfnamefont
  {W.~D.}\ \bibnamefont {Zhao}}, \bibinfo {author} {\bibfnamefont
  {L.}~\bibnamefont {He}}, \bibinfo {author} {\bibfnamefont {X.}~\bibnamefont
  {Zhang}}, \bibinfo {author} {\bibfnamefont {Z.~C.}\ \bibnamefont {Zhou}}, \
  and\ \bibinfo {author} {\bibfnamefont {L.~M.}\ \bibnamefont {Duan}},\ }\href
  {\doibase 10.1088/0256-307x/39/2/020502} {\bibfield  {journal} {\bibinfo
  {journal} {Chin. Phys. Lett.}\ }\textbf {\bibinfo {volume} {39}} (\bibinfo
  {year} {2022}),\ 10.1088/0256-307x/39/2/020502}\BibitemShut {NoStop}%
\bibitem [{\citenamefont {Chen}\ \emph {et~al.}(2021)\citenamefont {Chen},
  \citenamefont {Wu}, \citenamefont {Jiang}, \citenamefont {Lu}, \citenamefont
  {Peng},\ and\ \citenamefont {Du}}]{Chen2021NC}%
  \BibitemOpen
  \bibfield  {author} {\bibinfo {author} {\bibfnamefont {X.}~\bibnamefont
  {Chen}}, \bibinfo {author} {\bibfnamefont {Z.}~\bibnamefont {Wu}}, \bibinfo
  {author} {\bibfnamefont {M.}~\bibnamefont {Jiang}}, \bibinfo {author}
  {\bibfnamefont {X.~Y.}\ \bibnamefont {Lu}}, \bibinfo {author} {\bibfnamefont
  {X.}~\bibnamefont {Peng}}, \ and\ \bibinfo {author} {\bibfnamefont
  {J.}~\bibnamefont {Du}},\ }\href {\doibase 10.1038/s41467-021-26573-5}
  {\bibfield  {journal} {\bibinfo  {journal} {Nat. Commun.}\ }\textbf {\bibinfo
  {volume} {12}},\ \bibinfo {pages} {6281} (\bibinfo {year}
  {2021})}\BibitemShut {NoStop}%
\bibitem [{\citenamefont {Ashhab}\ and\ \citenamefont
  {Nori}(2010)}]{Ashhab2010PRA}%
  \BibitemOpen
  \bibfield  {author} {\bibinfo {author} {\bibfnamefont {S.}~\bibnamefont
  {Ashhab}}\ and\ \bibinfo {author} {\bibfnamefont {F.}~\bibnamefont {Nori}},\
  }\href {\doibase 10.1103/PhysRevA.81.042311} {\bibfield  {journal} {\bibinfo
  {journal} {Phys. Rev. A}\ }\textbf {\bibinfo {volume} {81}},\ \bibinfo
  {pages} {042311} (\bibinfo {year} {2010})}\BibitemShut {NoStop}%
\bibitem [{\citenamefont {Leroux}\ \emph {et~al.}(2017)\citenamefont {Leroux},
  \citenamefont {Govia},\ and\ \citenamefont {Clerk}}]{Leroux2017PRA}%
  \BibitemOpen
  \bibfield  {author} {\bibinfo {author} {\bibfnamefont {C.}~\bibnamefont
  {Leroux}}, \bibinfo {author} {\bibfnamefont {L.~C.~G.}\ \bibnamefont
  {Govia}}, \ and\ \bibinfo {author} {\bibfnamefont {A.~A.}\ \bibnamefont
  {Clerk}},\ }\href {\doibase 10.1103/PhysRevA.96.043834} {\bibfield  {journal}
  {\bibinfo  {journal} {Phys. Rev. A}\ }\textbf {\bibinfo {volume} {96}},\
  \bibinfo {pages} {043834} (\bibinfo {year} {2017})}\BibitemShut {NoStop}%
\bibitem [{\citenamefont {Chen}\ \emph {et~al.}(2020)\citenamefont {Chen},
  \citenamefont {Zhang}, \citenamefont {Fu},\ and\ \citenamefont
  {Zheng}}]{Chen2020PRA}%
  \BibitemOpen
  \bibfield  {author} {\bibinfo {author} {\bibfnamefont {X.-Y.}\ \bibnamefont
  {Chen}}, \bibinfo {author} {\bibfnamefont {Y.-Y.}\ \bibnamefont {Zhang}},
  \bibinfo {author} {\bibfnamefont {L.}~\bibnamefont {Fu}}, \ and\ \bibinfo
  {author} {\bibfnamefont {H.}~\bibnamefont {Zheng}},\ }\href {\doibase
  10.1103/PhysRevA.101.033827} {\bibfield  {journal} {\bibinfo  {journal}
  {Phys. Rev. A}\ }\textbf {\bibinfo {volume} {101}},\ \bibinfo {pages}
  {033827} (\bibinfo {year} {2020})}\BibitemShut {NoStop}%
\bibitem [{\citenamefont {Hacker}\ \emph {et~al.}(2019)\citenamefont {Hacker},
  \citenamefont {Welte}, \citenamefont {Daiss}, \citenamefont {Shaukat},
  \citenamefont {Ritter}, \citenamefont {Li},\ and\ \citenamefont
  {Rempe}}]{Hacker2019NP}%
  \BibitemOpen
  \bibfield  {author} {\bibinfo {author} {\bibfnamefont {B.}~\bibnamefont
  {Hacker}}, \bibinfo {author} {\bibfnamefont {S.}~\bibnamefont {Welte}},
  \bibinfo {author} {\bibfnamefont {S.}~\bibnamefont {Daiss}}, \bibinfo
  {author} {\bibfnamefont {A.}~\bibnamefont {Shaukat}}, \bibinfo {author}
  {\bibfnamefont {S.}~\bibnamefont {Ritter}}, \bibinfo {author} {\bibfnamefont
  {L.}~\bibnamefont {Li}}, \ and\ \bibinfo {author} {\bibfnamefont
  {G.}~\bibnamefont {Rempe}},\ }\href {\doibase 10.1038/s41566-018-0339-5}
  {\bibfield  {journal} {\bibinfo  {journal} {Nat. Photonics}\ }\textbf
  {\bibinfo {volume} {13}},\ \bibinfo {pages} {110} (\bibinfo {year}
  {2019})}\BibitemShut {NoStop}%
\bibitem [{\citenamefont {Bergmann}\ and\ \citenamefont {van
  Loock}(2016)}]{Bergmann2016PRA}%
  \BibitemOpen
  \bibfield  {author} {\bibinfo {author} {\bibfnamefont {M.}~\bibnamefont
  {Bergmann}}\ and\ \bibinfo {author} {\bibfnamefont {P.}~\bibnamefont {van
  Loock}},\ }\href {\doibase 10.1103/PhysRevA.94.042332} {\bibfield  {journal}
  {\bibinfo  {journal} {Phys. Rev. A}\ }\textbf {\bibinfo {volume} {94}},\
  \bibinfo {pages} {042332} (\bibinfo {year} {2016})}\BibitemShut {NoStop}%
\bibitem [{\citenamefont {Grimm}\ \emph {et~al.}(2020)\citenamefont {Grimm},
  \citenamefont {Frattini}, \citenamefont {Puri}, \citenamefont {Mundhada},
  \citenamefont {Touzard}, \citenamefont {Mirrahimi}, \citenamefont {Girvin},
  \citenamefont {Shankar},\ and\ \citenamefont {Devoret}}]{Grimm2020Nature}%
  \BibitemOpen
  \bibfield  {author} {\bibinfo {author} {\bibfnamefont {A.}~\bibnamefont
  {Grimm}}, \bibinfo {author} {\bibfnamefont {N.~E.}\ \bibnamefont {Frattini}},
  \bibinfo {author} {\bibfnamefont {S.}~\bibnamefont {Puri}}, \bibinfo {author}
  {\bibfnamefont {S.~O.}\ \bibnamefont {Mundhada}}, \bibinfo {author}
  {\bibfnamefont {S.}~\bibnamefont {Touzard}}, \bibinfo {author} {\bibfnamefont
  {M.}~\bibnamefont {Mirrahimi}}, \bibinfo {author} {\bibfnamefont {S.~M.}\
  \bibnamefont {Girvin}}, \bibinfo {author} {\bibfnamefont {S.}~\bibnamefont
  {Shankar}}, \ and\ \bibinfo {author} {\bibfnamefont {M.~H.}\ \bibnamefont
  {Devoret}},\ }\href {\doibase 10.1038/s41586-020-2587-z} {\bibfield
  {journal} {\bibinfo  {journal} {Nature}\ }\textbf {\bibinfo {volume} {584}},\
  \bibinfo {pages} {205} (\bibinfo {year} {2020})}\BibitemShut {NoStop}%
\bibitem [{\citenamefont {Li}\ \emph {et~al.}(2017)\citenamefont {Li},
  \citenamefont {Zou}, \citenamefont {Albert}, \citenamefont {Muralidharan},
  \citenamefont {Girvin},\ and\ \citenamefont {Jiang}}]{Li2017PRL}%
  \BibitemOpen
  \bibfield  {author} {\bibinfo {author} {\bibfnamefont {L.}~\bibnamefont
  {Li}}, \bibinfo {author} {\bibfnamefont {C.-L.}\ \bibnamefont {Zou}},
  \bibinfo {author} {\bibfnamefont {V.~V.}\ \bibnamefont {Albert}}, \bibinfo
  {author} {\bibfnamefont {S.}~\bibnamefont {Muralidharan}}, \bibinfo {author}
  {\bibfnamefont {S.~M.}\ \bibnamefont {Girvin}}, \ and\ \bibinfo {author}
  {\bibfnamefont {L.}~\bibnamefont {Jiang}},\ }\href {\doibase
  10.1103/PhysRevLett.119.030502} {\bibfield  {journal} {\bibinfo  {journal}
  {Phys. Rev. Lett.}\ }\textbf {\bibinfo {volume} {119}},\ \bibinfo {pages}
  {030502} (\bibinfo {year} {2017})}\BibitemShut {NoStop}%
\bibitem [{\citenamefont {Albert}\ \emph {et~al.}(2016)\citenamefont {Albert},
  \citenamefont {Shu}, \citenamefont {Krastanov}, \citenamefont {Shen},
  \citenamefont {Liu}, \citenamefont {Yang}, \citenamefont {Schoelkopf},
  \citenamefont {Mirrahimi}, \citenamefont {Devoret},\ and\ \citenamefont
  {Jiang}}]{Albert2016PRL}%
  \BibitemOpen
  \bibfield  {author} {\bibinfo {author} {\bibfnamefont {V.~V.}\ \bibnamefont
  {Albert}}, \bibinfo {author} {\bibfnamefont {C.}~\bibnamefont {Shu}},
  \bibinfo {author} {\bibfnamefont {S.}~\bibnamefont {Krastanov}}, \bibinfo
  {author} {\bibfnamefont {C.}~\bibnamefont {Shen}}, \bibinfo {author}
  {\bibfnamefont {R.-B.}\ \bibnamefont {Liu}}, \bibinfo {author} {\bibfnamefont
  {Z.-B.}\ \bibnamefont {Yang}}, \bibinfo {author} {\bibfnamefont {R.~J.}\
  \bibnamefont {Schoelkopf}}, \bibinfo {author} {\bibfnamefont
  {M.}~\bibnamefont {Mirrahimi}}, \bibinfo {author} {\bibfnamefont {M.~H.}\
  \bibnamefont {Devoret}}, \ and\ \bibinfo {author} {\bibfnamefont
  {L.}~\bibnamefont {Jiang}},\ }\href {\doibase 10.1103/PhysRevLett.116.140502}
  {\bibfield  {journal} {\bibinfo  {journal} {Phys. Rev. Lett.}\ }\textbf
  {\bibinfo {volume} {116}},\ \bibinfo {pages} {140502} (\bibinfo {year}
  {2016})}\BibitemShut {NoStop}%
\bibitem [{\citenamefont {Sun}\ \emph {et~al.}(2021)\citenamefont {Sun},
  \citenamefont {Zheng}, \citenamefont {Xiao}, \citenamefont {Gong},
  \citenamefont {He},\ and\ \citenamefont {Xia}}]{Sun2021PRL}%
  \BibitemOpen
  \bibfield  {author} {\bibinfo {author} {\bibfnamefont {F.-X.}\ \bibnamefont
  {Sun}}, \bibinfo {author} {\bibfnamefont {S.-S.}\ \bibnamefont {Zheng}},
  \bibinfo {author} {\bibfnamefont {Y.}~\bibnamefont {Xiao}}, \bibinfo {author}
  {\bibfnamefont {Q.}~\bibnamefont {Gong}}, \bibinfo {author} {\bibfnamefont
  {Q.}~\bibnamefont {He}}, \ and\ \bibinfo {author} {\bibfnamefont
  {K.}~\bibnamefont {Xia}},\ }\href {\doibase 10.1103/PhysRevLett.127.087203}
  {\bibfield  {journal} {\bibinfo  {journal} {Phys. Rev. Lett.}\ }\textbf
  {\bibinfo {volume} {127}},\ \bibinfo {pages} {087203} (\bibinfo {year}
  {2021})}\BibitemShut {NoStop}%
\bibitem [{\citenamefont {Joo}\ \emph {et~al.}(2011)\citenamefont {Joo},
  \citenamefont {Munro},\ and\ \citenamefont {Spiller}}]{Joo2011PRL}%
  \BibitemOpen
  \bibfield  {author} {\bibinfo {author} {\bibfnamefont {J.}~\bibnamefont
  {Joo}}, \bibinfo {author} {\bibfnamefont {W.~J.}\ \bibnamefont {Munro}}, \
  and\ \bibinfo {author} {\bibfnamefont {T.~P.}\ \bibnamefont {Spiller}},\
  }\href {\doibase 10.1103/PhysRevLett.107.083601} {\bibfield  {journal}
  {\bibinfo  {journal} {Phys. Rev. Lett.}\ }\textbf {\bibinfo {volume} {107}},\
  \bibinfo {pages} {083601} (\bibinfo {year} {2011})}\BibitemShut {NoStop}%
\bibitem [{\citenamefont {Hastrup}\ \emph {et~al.}(2021)\citenamefont
  {Hastrup}, \citenamefont {Park}, \citenamefont {Filip},\ and\ \citenamefont
  {Andersen}}]{Hastrup2021PRL}%
  \BibitemOpen
  \bibfield  {author} {\bibinfo {author} {\bibfnamefont {J.}~\bibnamefont
  {Hastrup}}, \bibinfo {author} {\bibfnamefont {K.}~\bibnamefont {Park}},
  \bibinfo {author} {\bibfnamefont {R.}~\bibnamefont {Filip}}, \ and\ \bibinfo
  {author} {\bibfnamefont {U.~L.}\ \bibnamefont {Andersen}},\ }\href {\doibase
  10.1103/PhysRevLett.126.153602} {\bibfield  {journal} {\bibinfo  {journal}
  {Phys. Rev. Lett.}\ }\textbf {\bibinfo {volume} {126}},\ \bibinfo {pages}
  {153602} (\bibinfo {year} {2021})}\BibitemShut {NoStop}%
\bibitem [{\citenamefont {S\'anchez Mu\~noz}\ and\ \citenamefont
  {Jaksch}(2021)}]{Sanchez2021PRL}%
  \BibitemOpen
  \bibfield  {author} {\bibinfo {author} {\bibfnamefont {C.}~\bibnamefont
  {S\'anchez Mu\~noz}}\ and\ \bibinfo {author} {\bibfnamefont {D.}~\bibnamefont
  {Jaksch}},\ }\href {\doibase 10.1103/PhysRevLett.127.183603} {\bibfield
  {journal} {\bibinfo  {journal} {Phys. Rev. Lett.}\ }\textbf {\bibinfo
  {volume} {127}},\ \bibinfo {pages} {183603} (\bibinfo {year}
  {2021})}\BibitemShut {NoStop}%
\bibitem [{\citenamefont {Xin}\ \emph {et~al.}(2021)\citenamefont {Xin},
  \citenamefont {Leong}, \citenamefont {Chen}, \citenamefont {Wang},\ and\
  \citenamefont {Lan}}]{Xin2021PRL}%
  \BibitemOpen
  \bibfield  {author} {\bibinfo {author} {\bibfnamefont {M.}~\bibnamefont
  {Xin}}, \bibinfo {author} {\bibfnamefont {W.~S.}\ \bibnamefont {Leong}},
  \bibinfo {author} {\bibfnamefont {Z.}~\bibnamefont {Chen}}, \bibinfo {author}
  {\bibfnamefont {Y.}~\bibnamefont {Wang}}, \ and\ \bibinfo {author}
  {\bibfnamefont {S.-Y.}\ \bibnamefont {Lan}},\ }\href {\doibase
  10.1103/PhysRevLett.127.183602} {\bibfield  {journal} {\bibinfo  {journal}
  {Phys. Rev. Lett.}\ }\textbf {\bibinfo {volume} {127}},\ \bibinfo {pages}
  {183602} (\bibinfo {year} {2021})}\BibitemShut {NoStop}%
\bibitem [{\citenamefont {Yang}\ and\ \citenamefont
  {Luo}(2022)}]{Yang2022CharacterizingSP}%
  \BibitemOpen
  \bibfield  {author} {\bibinfo {author} {\bibfnamefont {Y.-T.}\ \bibnamefont
  {Yang}}\ and\ \bibinfo {author} {\bibfnamefont {H.-G.}\ \bibnamefont {Luo}},\
  }\href@noop {} {\bibfield  {journal} {\bibinfo  {journal} {arXiv:2207.13285}\
  } (\bibinfo {year} {2022})}\BibitemShut {NoStop}%
\bibitem [{\citenamefont {Ying}\ \emph {et~al.}(2015)\citenamefont {Ying},
  \citenamefont {Liu}, \citenamefont {Luo}, \citenamefont {Lin},\ and\
  \citenamefont {You}}]{Ying2015PRA}%
  \BibitemOpen
  \bibfield  {author} {\bibinfo {author} {\bibfnamefont {Z.-J.}\ \bibnamefont
  {Ying}}, \bibinfo {author} {\bibfnamefont {M.}~\bibnamefont {Liu}}, \bibinfo
  {author} {\bibfnamefont {H.-G.}\ \bibnamefont {Luo}}, \bibinfo {author}
  {\bibfnamefont {H.-Q.}\ \bibnamefont {Lin}}, \ and\ \bibinfo {author}
  {\bibfnamefont {J.~Q.}\ \bibnamefont {You}},\ }\href {\doibase
  10.1103/PhysRevA.92.053823} {\bibfield  {journal} {\bibinfo  {journal} {Phys.
  Rev. A}\ }\textbf {\bibinfo {volume} {92}},\ \bibinfo {pages} {053823}
  (\bibinfo {year} {2015})}\BibitemShut {NoStop}%
\bibitem [{\citenamefont {Cong}\ \emph {et~al.}(2017)\citenamefont {Cong},
  \citenamefont {Sun}, \citenamefont {Liu}, \citenamefont {Ying},\ and\
  \citenamefont {Luo}}]{Cong2017PRA}%
  \BibitemOpen
  \bibfield  {author} {\bibinfo {author} {\bibfnamefont {L.}~\bibnamefont
  {Cong}}, \bibinfo {author} {\bibfnamefont {X.-M.}\ \bibnamefont {Sun}},
  \bibinfo {author} {\bibfnamefont {M.}~\bibnamefont {Liu}}, \bibinfo {author}
  {\bibfnamefont {Z.-J.}\ \bibnamefont {Ying}}, \ and\ \bibinfo {author}
  {\bibfnamefont {H.-G.}\ \bibnamefont {Luo}},\ }\href {\doibase
  10.1103/PhysRevA.95.063803} {\bibfield  {journal} {\bibinfo  {journal} {Phys.
  Rev. A}\ }\textbf {\bibinfo {volume} {95}},\ \bibinfo {pages} {063803}
  (\bibinfo {year} {2017})}\BibitemShut {NoStop}%
\bibitem [{\citenamefont {Cong}\ \emph {et~al.}(2019)\citenamefont {Cong},
  \citenamefont {Sun}, \citenamefont {Liu}, \citenamefont {Ying},\ and\
  \citenamefont {Luo}}]{Cong2019PRA}%
  \BibitemOpen
  \bibfield  {author} {\bibinfo {author} {\bibfnamefont {L.}~\bibnamefont
  {Cong}}, \bibinfo {author} {\bibfnamefont {X.-M.}\ \bibnamefont {Sun}},
  \bibinfo {author} {\bibfnamefont {M.}~\bibnamefont {Liu}}, \bibinfo {author}
  {\bibfnamefont {Z.-J.}\ \bibnamefont {Ying}}, \ and\ \bibinfo {author}
  {\bibfnamefont {H.-G.}\ \bibnamefont {Luo}},\ }\href {\doibase
  10.1103/PhysRevA.99.013815} {\bibfield  {journal} {\bibinfo  {journal} {Phys.
  Rev. A}\ }\textbf {\bibinfo {volume} {99}},\ \bibinfo {pages} {013815}
  (\bibinfo {year} {2019})}\BibitemShut {NoStop}%
\bibitem [{\citenamefont {Sun}\ \emph {et~al.}(2020)\citenamefont {Sun},
  \citenamefont {Cong}, \citenamefont {Eckle}, \citenamefont {Ying},\ and\
  \citenamefont {Luo}}]{Sun2020PRA}%
  \BibitemOpen
  \bibfield  {author} {\bibinfo {author} {\bibfnamefont {X.-M.}\ \bibnamefont
  {Sun}}, \bibinfo {author} {\bibfnamefont {L.}~\bibnamefont {Cong}}, \bibinfo
  {author} {\bibfnamefont {H.-P.}\ \bibnamefont {Eckle}}, \bibinfo {author}
  {\bibfnamefont {Z.-J.}\ \bibnamefont {Ying}}, \ and\ \bibinfo {author}
  {\bibfnamefont {H.-G.}\ \bibnamefont {Luo}},\ }\href {\doibase
  10.1103/PhysRevA.101.063832} {\bibfield  {journal} {\bibinfo  {journal}
  {Phys. Rev. A}\ }\textbf {\bibinfo {volume} {101}},\ \bibinfo {pages}
  {063832} (\bibinfo {year} {2020})}\BibitemShut {NoStop}%
\bibitem [{\citenamefont {Ying}(2022)}]{Ying2022QUTE}%
  \BibitemOpen
  \bibfield  {author} {\bibinfo {author} {\bibfnamefont {Z.}~\bibnamefont
  {Ying}},\ }\href {\doibase 10.1002/qute.202100165} {\bibfield  {journal}
  {\bibinfo  {journal} {Adv. Quantum Technol.}\ } (\bibinfo {year} {2022}),\
  10.1002/qute.202100165}\BibitemShut {NoStop}%
\bibitem [{\citenamefont {Chu}\ \emph {et~al.}(2021)\citenamefont {Chu},
  \citenamefont {Zhang}, \citenamefont {Yu},\ and\ \citenamefont
  {Cai}}]{Chu2021PRL}%
  \BibitemOpen
  \bibfield  {author} {\bibinfo {author} {\bibfnamefont {Y.}~\bibnamefont
  {Chu}}, \bibinfo {author} {\bibfnamefont {S.}~\bibnamefont {Zhang}}, \bibinfo
  {author} {\bibfnamefont {B.}~\bibnamefont {Yu}}, \ and\ \bibinfo {author}
  {\bibfnamefont {J.}~\bibnamefont {Cai}},\ }\href {\doibase
  10.1103/PhysRevLett.126.010502} {\bibfield  {journal} {\bibinfo  {journal}
  {Phys. Rev. Lett.}\ }\textbf {\bibinfo {volume} {126}},\ \bibinfo {pages}
  {010502} (\bibinfo {year} {2021})}\BibitemShut {NoStop}%
\bibitem [{\citenamefont {Ying}\ \emph {et~al.}(2022)\citenamefont {Ying},
  \citenamefont {Felicetti}, \citenamefont {Liu},\ and\ \citenamefont
  {Braak}}]{Ying2022Entropy}%
  \BibitemOpen
  \bibfield  {author} {\bibinfo {author} {\bibfnamefont {Z.~J.}\ \bibnamefont
  {Ying}}, \bibinfo {author} {\bibfnamefont {S.}~\bibnamefont {Felicetti}},
  \bibinfo {author} {\bibfnamefont {G.}~\bibnamefont {Liu}}, \ and\ \bibinfo
  {author} {\bibfnamefont {D.}~\bibnamefont {Braak}},\ }\href {\doibase
  10.3390/e24081015} {\bibfield  {journal} {\bibinfo  {journal} {Entropy}\
  }\textbf {\bibinfo {volume} {24}} (\bibinfo {year} {2022}),\
  10.3390/e24081015}\BibitemShut {NoStop}%
\bibitem [{\citenamefont {Casanova}\ \emph {et~al.}(2010)\citenamefont
  {Casanova}, \citenamefont {Romero}, \citenamefont {Lizuain}, \citenamefont
  {Garc\'{\i}a-Ripoll},\ and\ \citenamefont {Solano}}]{Casanova2010PRL}%
  \BibitemOpen
  \bibfield  {author} {\bibinfo {author} {\bibfnamefont {J.}~\bibnamefont
  {Casanova}}, \bibinfo {author} {\bibfnamefont {G.}~\bibnamefont {Romero}},
  \bibinfo {author} {\bibfnamefont {I.}~\bibnamefont {Lizuain}}, \bibinfo
  {author} {\bibfnamefont {J.~J.}\ \bibnamefont {Garc\'{\i}a-Ripoll}}, \ and\
  \bibinfo {author} {\bibfnamefont {E.}~\bibnamefont {Solano}},\ }\href
  {\doibase 10.1103/PhysRevLett.105.263603} {\bibfield  {journal} {\bibinfo
  {journal} {Phys. Rev. Lett.}\ }\textbf {\bibinfo {volume} {105}},\ \bibinfo
  {pages} {263603} (\bibinfo {year} {2010})}\BibitemShut {NoStop}%
\bibitem [{\citenamefont {Kenfack}\ and\ \citenamefont
  {\.{Z}yczkowski}(2004)}]{Kenfack2004JOB}%
  \BibitemOpen
  \bibfield  {author} {\bibinfo {author} {\bibfnamefont {A.}~\bibnamefont
  {Kenfack}}\ and\ \bibinfo {author} {\bibfnamefont {K.}~\bibnamefont
  {\.{Z}yczkowski}},\ }\href {\doibase 10.1088/1464-4266/6/10/003} {\bibfield
  {journal} {\bibinfo  {journal} {J. Opt. B: Quantum Semiclassical Opt.}\
  }\textbf {\bibinfo {volume} {6}},\ \bibinfo {pages} {396} (\bibinfo {year}
  {2004})}\BibitemShut {NoStop}%
\bibitem [{\citenamefont {Ashhab}(2013)}]{Ashhab2013PRA}%
  \BibitemOpen
  \bibfield  {author} {\bibinfo {author} {\bibfnamefont {S.}~\bibnamefont
  {Ashhab}},\ }\href {\doibase 10.1103/PhysRevA.87.013826} {\bibfield
  {journal} {\bibinfo  {journal} {Phys. Rev. A}\ }\textbf {\bibinfo {volume}
  {87}},\ \bibinfo {pages} {013826} (\bibinfo {year} {2013})}\BibitemShut
  {NoStop}%
\bibitem [{\citenamefont {Lin}\ \emph {et~al.}(2019)\citenamefont {Lin},
  \citenamefont {Liu}, \citenamefont {Chesi},\ and\ \citenamefont
  {Luo}}]{Lin2017JPCS}%
  \BibitemOpen
  \bibfield  {author} {\bibinfo {author} {\bibfnamefont {H.~Q.}\ \bibnamefont
  {Lin}}, \bibinfo {author} {\bibfnamefont {M.~X.}\ \bibnamefont {Liu}},
  \bibinfo {author} {\bibfnamefont {S.}~\bibnamefont {Chesi}}, \ and\ \bibinfo
  {author} {\bibfnamefont {H.~G.}\ \bibnamefont {Luo}},\ }\href {\doibase
  10.1088/1742-6596/1163/1/012003} {\bibfield  {journal} {\bibinfo  {journal}
  {J. Phys.: Conf. Ser.}\ }\textbf {\bibinfo {volume} {1163}},\ \bibinfo
  {pages} {012003} (\bibinfo {year} {2019})}\BibitemShut {NoStop}%
\bibitem [{\citenamefont {Gerry}\ \emph {et~al.}(2005)\citenamefont {Gerry},
  \citenamefont {Knight},\ and\ \citenamefont
  {Knight}}]{Gerry2005IntroductoryQO}%
  \BibitemOpen
  \bibfield  {author} {\bibinfo {author} {\bibfnamefont {C.}~\bibnamefont
  {Gerry}}, \bibinfo {author} {\bibfnamefont {P.}~\bibnamefont {Knight}}, \
  and\ \bibinfo {author} {\bibfnamefont {P.~L.}\ \bibnamefont {Knight}},\
  }\href@noop {} {\emph {\bibinfo {title} {Introductory quantum optics}}}\
  (\bibinfo  {publisher} {Cambridge university press},\ \bibinfo {year}
  {2005})\ p.\ \bibinfo {pages} {161}\BibitemShut {NoStop}%
\end{thebibliography}%

\end{document}